\documentclass[a4paper,10pt]{article}

\usepackage[utf8]{inputenc}
\usepackage[T1]{fontenc}

\usepackage{amsmath,amssymb,amsthm}

\usepackage{enumitem}

\usepackage{authblk}
\usepackage[a4paper,left=2.5cm,right=2.5cm,top=2cm,bottom=2cm]{geometry}

\usepackage{array}
\usepackage{booktabs}
\usepackage{multirow}
\usepackage{tabularx}
\newcolumntype{M}[1]{>{\vspace{3pt}\raggedleft\arraybackslash}m{#1}}

\usepackage{graphicx}
\usepackage[export]{adjustbox}
\usepackage{caption}
\usepackage{placeins}
\usepackage[labelformat=simple]{subcaption}

\usepackage{tikz}
\usetikzlibrary{
  3d,
  arrows,
  calc,
  decorations.markings,
  decorations.pathmorphing,
  decorations.pathreplacing,
  hobby,
  intersections,
  matrix,
  patterns,
  positioning,
  shapes,
  spy
}

\usepackage{pgfplots}
\usepackage{pgfplotstable}
\usepgfplotslibrary{
  external,
  fillbetween,
  groupplots,
  patchplots,
  polar,
  units
}
\pgfplotsset{
  compat=1.18,
  samples=1000,
  every axis/.append style={
    font=\large,
    line width=1pt,
    tick style={line width=0.8pt}
  }
}

\usepackage{algorithm}
\usepackage{algpseudocode}
\algtext*{EndWhile}
\algtext*{EndIf}
\algtext*{EndFor}
\algtext*{EndProcedure}

\usepackage{siunitx}
\usepackage{cite}

\usepackage{hyperref}
\usepackage{hypcap}
\hypersetup{
  colorlinks=false,
  allcolors=blue
}

\theoremstyle{plain}

\theoremstyle{remark}

\numberwithin{equation}{section}

\usepackage{bm}
\usepackage{dsfont}
\usepackage{upgreek}
\usepackage{stmaryrd}
\usepackage{color}
\usepackage{overpic}

\newcommand{\R}{\mathds{R}}

\newcommand{\C}{\mathds{C}}

\newcommand{\freeEnergy}{w}
\newcommand{\dissipot}{\psi}

\newcommand{\diff}{\mathop{}\!\mathrm{d}}

\newcommand{\Sym}[1]{\operatorname{Sym}(#1)}
\newcommand{\SymPlus}[1]{\operatorname{Sym}_+\left(#1\right)}

\DeclareMathOperator*{\argmin}{arg\,min}

\usepackage[normalem]{ulem}

\definecolor{resolvedgreen}{rgb}{0,0.55,0}

\newcommand{\feps}{\bm{\varepsilon}}
\newcommand{\fsigma}{\bm{\sigma}}

\newcommand{\fxi}{\bm{\xi}}

\newcommand{\dmax}{d_{\mathrm{max}}}
\newcommand{\ddotmax}{\dot d_{\mathrm{max}}}

\title{Scalar Entropic Damage Transport: Well-posed softening\\ with a single damage variable}

\author[1]{Jonas Nutz}
\author[1]{Vikram Laheri}
\author[1,2,3,*]{Matti Schneider}

\affil[1]{University of Duisburg-Essen, Institute of Engineering Mathematics, Essen, Germany}
\affil[2]{Center for Nanointegration Duisburg-Essen (CENIDE)}
\affil[3]{Fraunhofer Institute for Industrial Mathematics ITWM, Kaiserslautern, Germany}
\affil[*]{Correspondence to: \texttt{matti.schneider@uni-due.de}}

\date{}

\begin{document}

\maketitle

\begin{abstract}
\noindent
Entropic Damage Transport (EDT) represents material softening via a local continuum damage model while ensuring the well-posedness of the mechanical boundary value problem. It is based on a convexification of the time-incremental potential associated with a rate-limited scalar damage model. A one-dimensional distribution of damage states serves as the internal variable, and the number of these states grows with the number of time steps.\\
We show that the emergence of the damage distribution is an artifact of the time discretization. Surprisingly, the distribution concentrates at a single damage state in the time-continuous limit, and EDT converges to a rate-limited damage model with a scalar damage variable, whose free energy is jointly convex in the strain and the reciprocal degradation.\\
We exploit this hidden convexity by a frozen-metric discretization in terms of the reciprocal degradation, which renders every increment jointly convex. The resulting scalar update preserves damage irreversibility, the prescribed damage-rate limitation and thermomechanical consistency, reduces to a scalar equation and provides a symmetric and positive definite algorithmic tangent in closed form.\\
Computational experiments assess the accuracy and the efficiency of both formulations from the material point to the microstructure level, and demonstrate that the scalar update permits a finite-element computation of industrial size with more than $14$ million degrees of freedom.\\[\baselineskip]
\end{abstract}

\newpage
\section{Introduction}
\label{sec:intro}

\subsection{State of the art}
\label{sec:intro_state}

Local continuum damage mechanics describes the progressive loss of load-carrying capacity of a material by a degradation variable affecting the free energy. Its first concepts emerged from creep-rupture studies~\cite{Kachanov1999,Rabotnov1963} and were subsequently embedded into thermodynamically consistent constitutive frameworks~\cite{Lemaitre1985,SimoJu1987,Chaboche1988}, see the review by Krajcinovic~\cite{Krajcinovic1989} and the monograph by Murakami~\cite{Murakami2012}. Various variables are in use to carry the damage, among them scalar degradations and damage tensors~\cite{VoyiadjisKattan2009} as well as stiffness- or compliance-based variables~\cite{Ortiz1985,Krajcinovic1998}.

Softening renders such models useful for failure prediction~\cite{Lemaitre1986}, for instance of fiber-reinforced composites~\cite{PuckSchuermann2002,Rolland2016,HanhanSangid2021}. Computations which resolve the microstructure, be it of a lab-size coupon~\cite{Laheri2023} or of the unit cells of a multiscale chain~\cite{Spahn2014,Koebler2021,Magino2022}, typically evaluate a local constitutive law at every material point. In such a finite-element computation, however, damage tends to localize into a single band of elements whose width decreases with the element size. Thus, the material response depends on the mesh, and, without a regularization, the energy dissipated in the band typically tends to zero under refinement, as the dissipation per unit volume remains bounded whereas the volume of the band vanishes~\cite{DeBorst1993}. This deficiency of rate-independent softening models was recognized early~\cite{Bazant1976} and motivated nonlocal remedies~\cite{BazantBelytschkoChang1984}. For a comparison of these remedies, see Jir{\'a}sek~\cite{Jirasek1998}, and for a review, see Ba{\v{z}}ant and Jir{\'a}sek~\cite{Bazant2002}. Mathematically, softening typically deprives the incremental potential of a local damage model of its quasiconvexity, so that minimizers of the associated boundary value problem need not exist~\cite{Morrey1952,Dacorogna1982}. At the level of rates, localization may be accompanied by a loss of ellipticity of the boundary value problem for the displacement rate, which is detected when the acoustic tensor formed with the tangent stiffness ceases to be positive definite for some direction of propagation~\cite{Benallal1993}.

The remedies proposed in the literature fall into four classes.
\begin{itemize}
	\item \emph{Mesh adjustment.} Engineering practice commonly adapts the constitutive law to the mesh. The crack-band model~\cite{Bazant1983} ties the band width to a characteristic element length, which fixes the dissipated energy by a deliberate adjustment of the softening law to this length. Phenomenological approaches such as GISSMO~\cite{Andrade2016}, widespread in crashworthiness and sheet metal forming, reduce the mesh sensitivity by tabulating failure parameters over the element size. These parameters are tied to the discretization for which they were identified and have to be identified anew for every element size. The adjustment regularizes the dissipated energy, not the boundary value problem, whose solution still localizes into a single row of elements aligned with the mesh.
	\item \emph{Rate limitation.} This class starts from the observation that inertia bounds the speed at which a crack propagates~\cite{Freund1990} and thus how fast the stored energy may be dissipated. A quasi-static computation neglects inertia, so that nothing prevents the damage from jumping within a single time step, and the constitutive law has to supply this limit instead. Viscous models penalize a rapid damage evolution~\cite{Needleman1988,SluysDeBorst1992,Dube1996}, and the delay-damage models bound the damage rate by a characteristic time~\cite{Allix2003,Suffis2003,Allix2013}. In dynamics, this time and the elastic wave speed combine into an internal length~\cite{DeBorst1993}, whose regularizing effect fades for slow loading and may be absent for dynamic failure out of rest~\cite{Zghal2020}. Quasi-statically, the rate limitation bounds the damage increment per time step without introducing a length and is thus no spatial regularization by itself. Moreover, it generally alters the elastic domain, i.e., the range of driving forces under which the damage remains inactive, unless the threshold is adjusted accordingly. Without a term linear in the damage rate, the elastic domain is lost altogether, as the damage evolves under any positive driving force unless the free energy provides a threshold~\cite{JunkerBalzani2021,SJH}. Viscous regularization was also combined with relaxation~\cite{Junker2017,Langenfeld2018}, where the required viscosity may be unphysically high and its effect depends on the loading rate~\cite{Junker2022,Langenfeld2022}.
	\item \emph{Nonlocal and gradient-extended models.} If the damage at a material point depends on its neighborhood, a localization band cannot become narrower than an internal length, and the dissipated energy does not vanish under mesh refinement. This class comprises nonlocal averaging~\cite{PijaudierCabot1987,BazantPijaudier1988,LorentzAndrieux2003}, gradient-extended~\cite{Peerlings1996,Peerlings2001,Dimitrijevic2011} and micromorphic formulations~\cite{Forest2009,Brepols2017} as well as phase-field fracture~\cite{FrancfortMarigo1998,Miehe2010,WuNguyen2018}. Mathematically, the direct method requires coercivity and weak lower semicontinuity~\cite{Dacorogna1982}, and without quasiconvexity the latter fails, so that minimizing sequences oscillate on ever finer scales. Nonlocal and gradient terms upgrade weak to strong convergence of the damage field, by a compact Sobolev embedding or by the convolution with a fixed kernel. For a coercive functional, minimizers thus exist~\cite{MielkeRoubicek2015}, which, however, need not be unique or depend continuously on the load. In practice, the incremental energy of gradient-damage and phase-field models remains nonconvex jointly in the displacement and the damage, so that the solution of a time step may be nonunique and jump under a steady increase of the prescribed loading~\cite{PhamMarigoMaurini2011,FarrellMaurini2017}. Moreover, the internal length is typically not accessible to measurement~\cite{WuNguyen2018}, and if it approaches the size of the microstructure, first-order homogenization lacks the separation of scales it requires~\cite{DruganWillis1996,Gitman2007,Coenen2012}. Finally, the additional global field enlarges the system solved in every increment~\cite{Navidtehrani2021,Junker2022}.
	\item \emph{Relaxation.} A nonconvex incremental potential favors the formation of microstructures on ever finer scales~\cite{Carstensen2002}, which a finite-element computation resolves only down to the element size. Relaxation replaces the incremental potential by a generalized convex envelope -- the quasiconvex, the rank-one convex or the convex one~\cite{BalzaniOrtiz2012}. In contrast to nonlocal and gradient terms, it does not upgrade weak to strong convergence but renders the energy compatible with weak limits, as the quasiconvex envelope is weakly lower semicontinuous~\cite{Dacorogna1982}. Relaxation does not restore coercivity, however. As the degradation of a softening model tends to zero, the incremental energy loses its coercivity in the strain, and minimizing sequences need not be bounded. The envelope stands for a microstructure on a finer scale~\cite{KohnMueller1992,Carstensen2002,Miehe2004}, which relaxed damage models carry along in the evolution~\cite{Junker2017,KoehlerBalzani2023,Melchior2026}. No efficient algorithm is known for the quasiconvex envelope~\cite{Balzani2024,Lauff2026_OTModel}, one reason being that quasiconvexity lacks a local characterization~\cite{Kristensen1999}. Hence, the rank-one convex envelope~\cite{Balzani2024,Koehler2024} or the convex one~\cite{SJH,Lauff2026_OTModel} is taken instead, although rank-one convexity does not imply weak lower semicontinuity. Convexity alone ensures neither existence nor uniqueness. A lower semicontinuous and strongly convex incremental energy, however, admits a unique minimizer, which depends Lipschitz-continuously on the prescribed load~\cite{RockafellarWets1998}. Relaxation was also combined with a viscous regularization~\cite{Langenfeld2018}.
\end{itemize}

The work at hand is concerned with Damage Transport, which belongs to the fourth class but builds on the second. It originates from the work of Schwarz et al.~\cite{SJH}, who took up the bounded damage rate of the delay-damage models~\cite{Allix2003,Allix2013} to supply the coercivity which relaxation alone lacks, limited the damage increment of a time step accordingly and approximated the convex envelope of the resulting condensed energy numerically. Lauff et al.~\cite{Lauff2026_OTModel} computed this envelope in closed form, revealing that the internal variable becomes a distribution of damage states and the dissipation a transport cost between such distributions~\cite{Villani2009}. Subsequently, they introduced an entropic regularization, as is common in computational optimal transport~\cite{Cuturi2013,Peyre2015}, which renders the condensed incremental potential uniformly strongly convex~\cite{Lauff2026_OTRModel}. As a consequence, the incremental boundary value problem of small-strain elasticity is well-posed in the sense of Hadamard, i.e., a solution exists, is unique and depends continuously on the prescribed load.

For the resulting model, a fast material-point solver~\cite{Schneider2026FastOTR} and a redistribution strategy onto a fixed set of damage states at exact stiffness~\cite{Schneider2026_Redistribution} are available. The redistribution permits adaptive time steps and reduces the memory footprint of the implementation, at the price of an additional approximation of the distribution. Entropic Damage Transport thus combines a softening response with a local model, requires neither an additional field nor an internal length and leads to well-posed incremental boundary value problems, which renders it a promising approach. Its internal variable, however, remains a discrete distribution of damage states, no matter how efficiently it is updated.

\newpage
\subsection{Contributions}
\label{sec:intro_contributions}
Entropic Damage Transport was formulated for a fixed time discretization, and its damage distribution emerges from the time steps. We therefore asked ourselves whether more suitable discretizations exist, for instance ones which approximate the damage distribution by a fixed number of states.

To this end, we identified the continuous-time model which Entropic Damage Transport discretizes, and the answer is surprising: The effective degradation of Entropic Damage Transport converges to that of a scalar rate-limited damage model, and the damage distribution concentrates at the damage state of this model, with a mean square distance bounded by a multiple of the time step size, see Appendix~\ref{apx:continuous_limit} for the argument at material-point level. Hence, there is no need to approximate a distribution, and a scalar internal variable suffices, provided its time discretization retains the convexity of the increment.

To transfer the favorable properties of Damage Transport to a scalar damage variable, we reveal a hidden convex structure of the underlying scalar model. Used as the internal variable, the reciprocal degradation enters the damaged compliance linearly, and the free energy inherits the joint convexity of the Hooke energy in the strain and the compliance~\cite{Goerthofer2022}. The relaxed free energy of Damage Transport carries the linear average of this very quantity over the damage distribution~\cite{Lauff2026_OTModel,Lauff2026_OTRModel}. Thus, both free energies share the same convex structure. This joint convexity permits eliminating the damage distribution altogether, instead of updating it more efficiently~\cite{Schneider2026FastOTR,Schneider2026_Redistribution}.

However, in the scalar setting, the dissipation potential becomes state-dependent, as a metric converts the rate of the reciprocal degradation into the damage rate. As a result, the fully implicit increment loses its joint convexity beyond a critical time step size wherever the second derivative of the reciprocal degradation is positive.

We propose the \emph{frozen-metric discretization}, which ensures strong convexity of the condensed incremental potential. In more detail, it evaluates the metric at the previous state and adapts the semi-implicit schemes used for gradient systems with a state-dependent kinetic structure~\cite{Zaitzeff2021,Mielke2023} to the derivative of the reciprocal degradation with respect to the damage. The frozen-metric discretization is a time-discrete model in its own right, which shares the continuous-time limit of Entropic Damage Transport at material-point level but differs from Entropic Damage Transport at finite time step sizes. The scalar model underlying Entropic Damage Transport may be considered as a delay-damage model~\cite{Allix2003,Allix2013} whose damage-rate limitation is enforced by a logarithmic barrier. The key novelty lies in its internal variable, which reveals the hidden convex structure, and in its time discretization, which preserves this structure. In particular, our findings reveal that Entropic Damage Transport belongs, in fact, to the second class of mitigation strategies. Its well-posedness stems from the damage-rate limitation, combined with a suitable internal variable and time discretization, and not from the relaxation. In retrospect, the relaxation of Schwarz et al.~\cite{SJH} was dispensable, as a more suitable time discretization of the rate-limited model would have sufficed.

The designed model offers the following advantages.
\begin{itemize}
	\item Each material point carries a single scalar internal variable, the reciprocal degradation.
	\item For an initial damage distribution concentrated at a single state, the effective degradation of Entropic Damage Transport and the frozen-metric states converge to one and the same scalar evolution at material-point level as the time step size vanishes, see Appendix~\ref{apx:convergence}.
	\item The increment is jointly convex in the strain and the reciprocal degradation for every positive time step size. The consistent algorithmic material tangent is symmetric and uniformly positive definite. Moreover, it is available in closed form.
	\item The admissible states of one increment form an interval known in closed form. The update preserves damage irreversibility and the prescribed damage-rate limitation, and it satisfies the reduced discrete Clausius--Duhem inequality. Computing the update requires solving a single scalar equation, for which the undamped Newton method converges monotonically.
\end{itemize}
In Sec.~\ref{sec:theory_scalar_model}, we describe the scalar model underlying Damage Transport, Sec.~\ref{sec:theory_hidden_convexity} introduces the reciprocal degradation as the internal variable and Sec.~\ref{sec:theory_frozen_metric} the frozen-metric increment. Sec.~\ref{sec:numerics} compares both formulations at material-point level, on a fiber-reinforced unit cell and on a notched specimen, and applies the scalar update to a jet-engine bracket with more than $14$ million degrees of freedom. In Appendix~\ref{apx:convergence}, we provide the convergence arguments at material-point level, and Appendix~\ref{apx:frozen_metric_properties} collects the arguments for the properties of the frozen-metric update.

\section{Hidden convexity and frozen-metric time discretization}
\label{sec:theory}

\subsection{The scalar model underlying Damage Transport}
\label{sec:theory_scalar_model}

We consider the rate-limited scalar damage model underlying the relaxation approach by Schwarz et al.~\cite{SJH} and the subsequent Damage-Transport framework~\cite{Lauff2026_OTModel,Lauff2026_OTRModel}. The model is formulated as a generalized standard material~\cite{Halphen1975} in an isothermal small-strain setting. Let
\begin{equation}\label{eq:theory_scalar_model_undamaged_energy}
	\freeEnergy_0(\feps) = \frac{1}{2}\feps:\C:\feps, \quad \feps\in\Sym{3},
\end{equation}
denote the undamaged elastic energy, where $\Sym{3}$ stands for the space of symmetric second-order tensors in three dimensions and the elasticity tensor $\C:\Sym{3}\rightarrow\Sym{3}$ is symmetric and positive definite. Let the scalar damage variable take values in the interval
\begin{equation}\label{eq:theory_scalar_model_damage_space}
	d\in\mathcal{D}:=[0,\dmax), \qquad 0<\dmax\leq\infty ,
\end{equation}
where the maximum damage $\dmax$ may be finite or infinite.
We consider a given degradation function
\begin{equation}\label{eq:theory_scalar_model_degradation}
	f:\mathcal{D}\rightarrow(0,1]
\end{equation}
which is continuous and strictly monotonically decreasing and satisfies the conditions
\begin{equation}\label{eq:theory_scalar_model_degradation_boundary}
	f(0)=1 \quad\text{and}\quad	\lim_{d\rightarrow\dmax}f(d)=0.
\end{equation}
Following the structural assumptions of Damage Transport~\cite{Lauff2026_OTModel,Lauff2026_OTRModel}, we introduce the reciprocal degradation
\begin{equation}\label{eq:theory_scalar_model_reciprocal_degradation}
	g:\mathcal{D}\rightarrow[1,\infty),	\qquad g(d)=\frac{1}{f(d)},
\end{equation}
and assume that the function $g$ is strictly convex. The conditions~\eqref{eq:theory_scalar_model_degradation_boundary} directly imply
\begin{equation}\label{eq:theory_scalar_model_reciprocal_degradation_boundary}
	g(0)=1 \quad\text{and}\quad	\lim_{d\rightarrow\dmax}g(d)=\infty.
\end{equation}
As the reciprocal of the degradation function $f$, which is positive for all damage values, the function $g$ is continuous, strictly increasing and, by the boundary conditions~\eqref{eq:theory_scalar_model_reciprocal_degradation_boundary}, a bijection from the interval $\mathcal{D}$ onto the interval $[1,\infty)$. For the differential arguments below, we additionally suppose that the conditions
\begin{equation}\label{eq:theory_scalar_model_regularities}
	g\in C^2(\mathcal{D}), \qquad g'(d)>0 \quad \text{and} \quad g''(d)\geq 0, \quad d\in\mathcal{D},
\end{equation}
hold. A prominent example satisfying the stated assumptions on an unbounded damage interval is the exponential degradation~\cite{DimitrijevicHackl2008}
\begin{equation} \label{eq:theory_scalar_model_exponential_degradation}
	f(d)=\exp(-d) \quad\text{and}\quad g(d)=\exp(d), \qquad d\in[0,\infty).
\end{equation}
With the undamaged energy~\eqref{eq:theory_scalar_model_undamaged_energy} and the degradation function~\eqref{eq:theory_scalar_model_degradation}, the free energy is defined by
\begin{equation}\label{eq:theory_scalar_model_free_energy}
	\freeEnergy(\feps,d) := f(d)\,\freeEnergy_0(\feps) = \frac{\freeEnergy_0(\feps)}{g(d)}, \quad \feps\in\Sym{3} \quad\text{and}\quad d\in\mathcal{D}.
\end{equation}
The corresponding stress computes as
\begin{equation}\label{eq:theory_scalar_model_stress}
	\fsigma	:= \frac{\partial\freeEnergy}{\partial\feps}(\feps,d) = f(d) \, \C:\feps.
\end{equation}
The damage rate is restricted by a constitutive damage-rate limitation $\ddotmax$. With the unregularized damage-initiation threshold $r$ and the entropic regularization parameter $\lambda$ intrinsic to Entropic Damage Transport~\cite{Lauff2026_OTRModel}, subject to the restrictions
\begin{equation}\label{eq:theory_scalar_model_dissipation_parameters}
	\ddotmax>0, \qquad r\geq 0 \quad\text{and}\quad \lambda>0 ,
\end{equation}
we consider the dissipation potential of the damage rate $\dot d$
\begin{equation}\label{eq:theory_scalar_model_dissipation}
	\dissipot (\dot{d})	:=
	\begin{cases}
		r \, \dot{d} - \lambda \ddotmax \log\left(1-\dot{d}/\ddotmax\right) , & 0 \leq \dot{d} < \ddotmax,\\
		+\infty, & \text{otherwise}.
	\end{cases}
\end{equation}
The dissipation potential~\eqref{eq:theory_scalar_model_dissipation} is a barrier version of the bounded-rate concept~\cite{Allix2013,SJH}. Here, the damage-rate limitation $\ddotmax$ carries the unit of an inverse time, whereas the threshold $r$ and the parameter $\lambda$ are energy densities, i.e., they carry the unit of a stress.
For a damage rate below the damage-rate limitation, differentiating the dissipation potential~\eqref{eq:theory_scalar_model_dissipation} yields the identities
\begin{equation}\label{eq:theory_scalar_model_dissipation_derivatives}
	\dissipot'(\dot{d}) =	r + \frac{\lambda}{1 - \dot{d}/\ddotmax} \quad \text{and} \quad	\dissipot''(\dot{d}) = \frac{\lambda/\ddotmax}{(1 - \dot{d}/\ddotmax)^2}>0.
\end{equation}
At a vanishing damage rate, both identities hold for the right-sided derivatives. Thus, the dissipation potential is strictly convex on its effective domain. Moreover, the logarithmic term diverges as the damage rate $\dot{d}$ approaches the damage-rate limitation $\ddotmax$ and therefore prevents the upper damage-rate bound from being attained. The thermodynamic force associated with the damage variable reads
\begin{equation}\label{eq:theory_scalar_model_driving_force}
	Y(\feps,d) := -\frac{\partial\freeEnergy}{\partial d}(\feps,d) = -f'(d) \, \freeEnergy_0(\feps)	= \frac{g'(d)}{g(d)^2} \, \freeEnergy_0(\feps).
\end{equation}
The evolution of the damage variable is governed by Biot's equation~\cite{Biot1965}
\begin{equation}\label{eq:theory_scalar_model_Biot}
	Y(\feps,d)\in\partial\dissipot(\dot d),
\end{equation}
where $\partial\dissipot$ denotes the subdifferential of the convex dissipation potential~\eqref{eq:theory_scalar_model_dissipation}. On the active branch, inserting the first relation in Eq.~\eqref{eq:theory_scalar_model_dissipation_derivatives} into Biot's equation~\eqref{eq:theory_scalar_model_Biot} yields the result
\begin{equation}\label{eq:theory_scalar_model_Biot_active}
	Y(\feps,d) = r + \frac{\lambda}{1-\dot d/\ddotmax}.
\end{equation}
At a vanishing damage rate, the subdifferential of the dissipation potential~\eqref{eq:theory_scalar_model_dissipation} consists of its right derivative together with the normal cone of the effective domain at its lower endpoint, which adds every smaller value,
\begin{equation}\label{eq:theory_scalar_model_subdifferential_zero}
	\partial\dissipot(0) = (-\infty,\, r+\lambda] .
\end{equation}
Thus, the solution of Biot's equation~\eqref{eq:theory_scalar_model_Biot}, obtained by solving Eq.~\eqref{eq:theory_scalar_model_Biot_active} for the damage rate and accounting for the subdifferential~\eqref{eq:theory_scalar_model_subdifferential_zero}, is the closed-form evolution law
\begin{equation}\label{eq:theory_scalar_model_evolution}
	\dot d = \ddotmax \frac{\left\langle Y(\feps,d)-r-\lambda\right\rangle_+} {Y(\feps,d)-r},
\end{equation}
where $\langle\cdot\rangle_+$ denotes the positive part and the fraction is defined to be zero whenever its numerator vanishes. Together with the definition~\eqref{eq:theory_scalar_model_driving_force}, the right-hand side of Eq.~\eqref{eq:theory_scalar_model_evolution} depends only on the current strain and damage state. In particular, damage remains inactive as long as the driving force does not exceed the total threshold,
\begin{equation}\label{eq:theory_scalar_model_elastic_domain}
	Y(\feps,d)\leq r+\lambda,
\end{equation}
whereas the inequalities
\begin{equation}\label{eq:theory_scalar_model_rate_bound}
	0<\dot d<\ddotmax
\end{equation}
hold on the active branch. The sum $r+\lambda$ is thus the driving force at which damage initiates, and the parameter $\lambda$ sets the scale of the overstress.

The scalar evolution~\eqref{eq:theory_scalar_model_evolution} constitutes the continuous rate-limited damage model underlying the time-discrete Entropic Damage-Transport model by Lauff et al.~\cite{Lauff2026_OTRModel}. At material-point level and for an initial damage distribution concentrated at a single damage state, the effective degradation of Entropic Damage Transport converges to the degradation along the scalar evolution~\eqref{eq:theory_scalar_model_evolution} as the time step size tends to zero, see Appendix~\ref{apx:continuous_limit}.

\subsection{Reciprocal degradation and hidden convexity}
\label{sec:theory_hidden_convexity}

The free energy~\eqref{eq:theory_scalar_model_free_energy} is not jointly convex in the strain and the damage variable, as its Hessian is indefinite wherever the strain does not vanish and the second derivative of the reciprocal degradation is positive. On the other hand, the Hooke energy
\begin{equation}\label{eq:theory_hidden_convexity_Hooke_energy}
	\freeEnergy^{\mathrm{Hooke}}: \Sym{3} \times \SymPlus{\Sym{3}} \rightarrow \R_{\geq 0}, \quad (\feps, \mathbb{S}) \mapsto \frac{1}{2} \feps : \mathbb{S}^{-1}:\feps,
\end{equation}
where $\SymPlus{\Sym{3}}$ denotes the symmetric and positive definite linear maps on $\Sym{3}$, is jointly convex in the strain and the compliance tensor~\cite{Goerthofer2022}, as it is a matrix-fractional function~\cite{BoydVandenberghe2004}. In the scalar damage model at hand, the damaged compliance tensor takes the form
\begin{equation} \label{eq:theory_hidden_convexity_compliance}
	\mathbb{S}(d) := \frac{1}{f(d)} \, \C^{-1} = g(d)\,\C^{-1}, \qquad d\in\mathcal{D}.
\end{equation}
Thus, the reciprocal degradation value $g(d)$ acts as a scalar compliance variable. The same structure underlies the relaxed free energy of Damage Transport~\cite{Lauff2026_OTModel,Lauff2026_OTRModel}, where the effective reciprocal degradation is obtained as a linear average over the damage distribution, so that a damage distribution stores the free energy of the single reciprocal degradation value at this average, see Fig.~\ref{fig:apx_continuous_limit_distribution} in Appendix~\ref{apx:continuous_limit}. These observations suggest using the reciprocal degradation value~\eqref{eq:theory_scalar_model_reciprocal_degradation} directly as the scalar internal variable, i.e., we define
\begin{equation} \label{eq:theory_hidden_convexity_transformation}
	z:=g(d), \qquad d\in\mathcal{D}.
\end{equation}
As the reciprocal degradation is a bijection onto the interval $[1,\infty)$, the inverse function
\begin{equation} \label{eq:theory_hidden_convexity_inverse}
	g^{-1}:[1,\infty)\rightarrow\mathcal{D}
\end{equation}
is well-defined, and the original damage variable may be recovered from the reciprocal degradation variable
\begin{equation}\label{eq:theory_hidden_convexity_inverse_relation}
	d = g^{-1}(z) \quad \text{for} \quad z \in [1,\infty).
\end{equation}
By the inverse function theorem, the inverse function $g^{-1}$ is twice continuously differentiable, as the reciprocal degradation $g$ is twice continuously differentiable with a positive derivative, see Eq.~\eqref{eq:theory_scalar_model_regularities}. Differentiating the identity
\begin{equation}\label{eq:theory_hidden_convexity_inverse_identity}
	g(g^{-1}(z)) = z, \qquad z\in[1,\infty),
\end{equation}
twice, the chain rule contributing a further factor $(g^{-1})'$ in the second step, yields the identities
\begin{equation} \label{eq:theory_hidden_convexity_inverse_derivatives}
	(g^{-1})'(z) = \frac{1}{g'(g^{-1}(z))} >0 \quad\text{and}\quad (g^{-1})''(z) = -\frac{g''(g^{-1}(z))} {[g'(g^{-1}(z))]^3} \leq0, \qquad z\in[1,\infty).
\end{equation}
Thus, the inverse function $g^{-1}$ is strictly monotonically increasing and, by the convexity assumption~\eqref{eq:theory_scalar_model_regularities}, concave. In terms of the reciprocal degradation variable defined in Eq.~\eqref{eq:theory_hidden_convexity_transformation}, the free energy~\eqref{eq:theory_scalar_model_free_energy} takes the form
\begin{equation}\label{eq:theory_hidden_convexity_transformed_energy}
	\widehat{\freeEnergy}(\feps,z) = \frac{1}{2z} \, \feps:\C:\feps \quad \text{for} \quad	\feps\in\Sym{3} \quad\text{and}\quad z\in[1,\infty).
\end{equation}
The transformed free energy~\eqref{eq:theory_hidden_convexity_transformed_energy} corresponds to the Hooke energy~\eqref{eq:theory_hidden_convexity_Hooke_energy} restricted to compliance tensors of the form
\begin{equation}\label{eq:theory_hidden_convexity_transformed_compliance}
	\widehat{\mathbb{S}}(z):=z \, \C^{-1}, \qquad z\in[1,\infty).
\end{equation}
As the compliance~\eqref{eq:theory_hidden_convexity_transformed_compliance} depends linearly on the reciprocal degradation variable, the transformed free energy is the composition of the Hooke energy with an affine map, and the joint convexity of the Hooke energy in the strain and the compliance tensor~\cite{Goerthofer2022} therefore carries over to the variables $\feps$ and $z$. The free energy thus becomes jointly convex under a change of variables, as in the hidden convexity of optimization problems that turn convex after a nonlinear transformation~\cite{BenTalTeboulle1996,Xia2020}. The joint convexity is not strict, because the transformed free energy is positively homogeneous of degree one in both variables jointly and thus linear along every ray from the origin, within its domain.
To express the dissipation potential~\eqref{eq:theory_scalar_model_dissipation} in terms of the reciprocal degradation variable, we differentiate Eq.~\eqref{eq:theory_hidden_convexity_transformation} with respect to time and obtain the expression
\begin{equation}\label{eq:theory_hidden_convexity_rate_transformation_preliminary}
	\dot{z} = g'(d) \, \dot{d}.
\end{equation}
Taking into account the inverse relation~\eqref{eq:theory_hidden_convexity_inverse_relation}, Eq.~\eqref{eq:theory_hidden_convexity_rate_transformation_preliminary} is equivalent to the formula
\begin{equation}\label{eq:theory_hidden_convexity_rate_transformation}
	\dot{d} = \frac{\dot{z}}{m(z)}, \quad z \in [1,\infty),
\end{equation}
where we introduce the \emph{metric}, which converts the rate of the reciprocal degradation variable into the damage rate,
\begin{equation}\label{eq:theory_hidden_convexity_metric}
	m:[1,\infty) \rightarrow (0,\infty), \qquad z\mapsto m(z):=g'(g^{-1}(z)) = \frac{1}{(g^{-1})'(z)}.
\end{equation}
The function $m$ takes only positive values, because the derivative of the reciprocal degradation is positive by assumption~\eqref{eq:theory_scalar_model_regularities}. Thus, the original dissipation potential~\eqref{eq:theory_scalar_model_dissipation} induces the state-dependent dissipation potential
\begin{equation}\label{eq:theory_hidden_convexity_transformed_dissipation}
	\widehat{\dissipot} : [1,\infty) \times \R \rightarrow [0,+\infty], \quad \widehat{\dissipot}(z,\dot{z}) = \dissipot\left(\frac{\dot{z}}{m(z)}\right).
\end{equation}
Taking into account the domain of the original dissipation potential~\eqref{eq:theory_scalar_model_dissipation}, the transformed dissipation potential~\eqref{eq:theory_hidden_convexity_transformed_dissipation} is finite under the condition
\begin{equation} \label{eq:theory_hidden_convexity_transformed_dissipation_domain}
	0 \leq \dot{z} < m(z)\,\ddotmax \quad \text{for} \quad z \in [1,\infty).
\end{equation}
For every fixed state, the rate transformation~\eqref{eq:theory_hidden_convexity_rate_transformation} is linear with a positive factor. Thus, the function $\widehat{\dissipot}(z,\cdot)$ inherits the strict convexity of the original dissipation potential $\dissipot$. This convexity property concerns the rate variable \emph{for a fixed state} only. In general, the transformed dissipation potential is not jointly convex in the variables $z$ and $\dot z$, as the metric $m$ depends on the state.

As the coordinate transformation~\eqref{eq:theory_hidden_convexity_transformation} is invertible, the transformed free energy~\eqref{eq:theory_hidden_convexity_transformed_energy} and the transformed dissipation potential~\eqref{eq:theory_hidden_convexity_transformed_dissipation} describe the same generalized standard material in terms of the reciprocal degradation variable $z$. Indeed, the driving force of the reciprocal degradation variable,
\begin{equation}\label{eq:theory_hidden_convexity_transformed_driving_force}
	\widehat Y(\feps,z) := -\frac{\partial\widehat{\freeEnergy}}{\partial z}(\feps,z) = \frac{\freeEnergy_0(\feps)}{z^2},
\end{equation}
and the subdifferential of the transformed dissipation potential with respect to the rate are related to the driving force~\eqref{eq:theory_scalar_model_driving_force} and to the subdifferential of the original dissipation potential by the identities
\begin{equation}\label{eq:theory_hidden_convexity_force_relations}
	m(z)\,\widehat Y(\feps,z) = Y\big(\feps,g^{-1}(z)\big) \quad\text{and}\quad \partial_{\dot z}\widehat{\dissipot}(z,\dot z) = \frac{1}{m(z)}\,\partial\dissipot\big(\dot z/m(z)\big).
\end{equation}
Thus, Biot's equation~\eqref{eq:theory_scalar_model_Biot} is equivalent to
\begin{equation}\label{eq:theory_hidden_convexity_transformed_Biot}
	\widehat Y(\feps,z)\in\partial_{\dot z}\widehat{\dissipot}(z,\dot z).
\end{equation}
For the exponential degradation~\eqref{eq:theory_scalar_model_exponential_degradation}, the coordinate transformation~\eqref{eq:theory_hidden_convexity_transformation} and the inverse relation~\eqref{eq:theory_hidden_convexity_inverse_relation} specialize to
\begin{equation}\label{eq:theory_hidden_convexity_exponential_transformation}
	z=\exp(d), \quad \text{i.e.,} \quad d = \log z \quad \text{for} \quad z \in [1,\infty).
\end{equation}
Taking into account the definition~\eqref{eq:theory_hidden_convexity_metric}, we obtain the relation
\begin{equation}\label{eq:theory_hidden_convexity_exponential_metric}
	m(z)=z, \qquad z\in[1,\infty),
\end{equation}
and, consequently, the rate transformation~\eqref{eq:theory_hidden_convexity_rate_transformation} reads
\begin{equation}\label{eq:theory_hidden_convexity_exponential_kinetics}
	\dot d=\frac{\dot z}{z}.
\end{equation}
The transformed dissipation potential~\eqref{eq:theory_hidden_convexity_transformed_dissipation} therefore takes the simple form
\begin{equation}\label{eq:theory_hidden_convexity_exponential_dissipation}
	\widehat{\dissipot}(z,\dot z) = \dissipot\left(\frac{\dot z}{z}\right), \qquad z\in[1,\infty).
\end{equation}
This potential is not jointly convex in the reciprocal degradation and its rate, as its Hessian has a negative determinant at every positive rate.

\subsection{Frozen-metric time discretization}
\label{sec:theory_frozen_metric}

We consider a discretization in time with a positive time step size $\triangle t$ and suppose that the previous reciprocal degradation state $z_n$ is given, and we write $z$ for the current state $z_{n+1}$. A fully implicit Euler discretization of the generalized standard material described by the transformed free energy~\eqref{eq:theory_hidden_convexity_transformed_energy} and the transformed dissipation potential~\eqref{eq:theory_hidden_convexity_transformed_dissipation} would be governed by the incremental potential
\begin{equation}\label{eq:theory_frozen_metric_fully_implicit_potential}
	\freeEnergy_{\mathrm{inc}}^{\mathrm{full}}(\feps,z \,| z_n) := \widehat{\freeEnergy}(\feps,z) + \triangle t\,\widehat{\dissipot}\left(z, \frac{z - z_n}{\triangle t}\right).
\end{equation}
Taking into account Eqs.~\eqref{eq:theory_hidden_convexity_transformed_energy} and~\eqref{eq:theory_hidden_convexity_transformed_dissipation}, the incremental potential~\eqref{eq:theory_frozen_metric_fully_implicit_potential} takes the explicit form
\begin{equation}\label{eq:theory_frozen_metric_fully_implicit_potential_explicit}
	\freeEnergy_{\mathrm{inc}}^{\mathrm{full}}(\feps,z\,| z_n) = \frac{1}{2 z} \, \feps:\C:\feps + \triangle t\,\dissipot\left(\frac{z - z_n}{\triangle t \, m(z)}\right).
\end{equation}
Although the elastic contribution in Eq.~\eqref{eq:theory_frozen_metric_fully_implicit_potential_explicit} is jointly convex in the variables $\feps$ and $z$, the argument of the dissipation potential depends nonlinearly on the current state through the function $m(z)$. Thus, the convexity of the transformed free energy and the separate convexity of the transformed dissipation potential in the rate variable do not imply joint convexity of the fully implicit incremental potential. For the exponential degradation~\eqref{eq:theory_scalar_model_exponential_degradation}, the metric~\eqref{eq:theory_hidden_convexity_exponential_metric} leads to the explicit expression
\begin{equation}\label{eq:theory_frozen_metric_fully_implicit_exponential}
	\freeEnergy_{\mathrm{inc}}^{\mathrm{full}}(\feps,z\,| z_n) = \frac{1}{2 z} \, \feps:\C:\feps + \triangle t\,\dissipot\left(\frac{z - z_n}{\triangle t \, z}\right).
\end{equation}
Joint convexity would in particular require the restriction of the incremental potential~\eqref{eq:theory_frozen_metric_fully_implicit_exponential} to a vanishing strain, i.e., its dissipation contribution alone, to be convex in the current state. Differentiating that contribution twice at the previous state and taking into account the identities~\eqref{eq:theory_scalar_model_dissipation_derivatives} yields
\begin{equation}\label{eq:theory_frozen_metric_fully_implicit_curvature}
	\left. \frac{\diff^2}{\diff z^2} \, \triangle t\,\dissipot\!\left(\frac{z-z_n}{\triangle t\,z}\right) \right|_{z=z_n} = \frac{1}{z_n^2} \left[ \frac{\lambda}{\triangle t\,\ddotmax} - 2(r+\lambda) \right] ,
\end{equation}
which is negative whenever the time step size exceeds the threshold
\begin{equation}\label{eq:theory_frozen_metric_fully_implicit_threshold}
	\triangle t > \frac{\lambda}{2\,\ddotmax\,(r+\lambda)} .
\end{equation}
The fully implicit increment~\eqref{eq:theory_frozen_metric_fully_implicit_exponential} therefore loses joint convexity beyond the threshold~\eqref{eq:theory_frozen_metric_fully_implicit_threshold}. For a general reciprocal degradation, the same computation with the derivative $m'$ of the metric yields the curvature
\begin{equation}\label{eq:theory_frozen_metric_fully_implicit_curvature_general}
	\frac{1}{[m(z_n)]^2} \left[ \frac{\lambda}{\triangle t\,\ddotmax} - 2(r+\lambda)\,m'(z_n) \right] \quad\text{with}\quad m'(z_n) = \frac{g''(d_n)}{g'(d_n)} \quad\text{and}\quad d_n := g^{-1}(z_n),
\end{equation}
so that the loss of joint convexity beyond a finite time step size is not special to the exponential degradation but occurs wherever the second derivative of the reciprocal degradation is positive. 

Here, the underlying mechanism is the following. The argument of the dissipation potential depends concavely on the current state, and this dependence contributes negatively to the curvature. The strictly convex dissipation potential compensates this contribution only for sufficiently small time step sizes. Gradient systems with state-dependent kinetic structures are commonly approximated by semi-implicit schemes which evaluate selected state-dependent quantities at the previous time step. Mielke~\cite[Sec.~3.2]{Mielke2023} evaluates the dissipation potential at the previous state, and Zaitzeff et al.~\cite{Zaitzeff2021} construct energy-stable semi-implicit Runge--Kutta schemes of high order. Following this strategy, we evaluate the state-dependent dissipation potential at the reciprocal degradation state $z_n$. We refer to the resulting approximation as a \emph{frozen-metric discretization}. The corresponding incremental potential is defined by
\begin{equation}\label{eq:theory_frozen_metric_incremental_potential}
	\freeEnergy_{\mathrm{inc}}(\feps,z\,| z_n) := \widehat{\freeEnergy}(\feps,z) + \triangle t\,\widehat{\dissipot}\left(z_n, \frac{z - z_n}{\triangle t}\right).
\end{equation}
Taking into account Eqs.~\eqref{eq:theory_hidden_convexity_transformed_energy} and~\eqref{eq:theory_hidden_convexity_transformed_dissipation}, the incremental potential~\eqref{eq:theory_frozen_metric_incremental_potential} admits the equivalent representation
\begin{equation} \label{eq:theory_frozen_metric_incremental_potential_explicit}
	\freeEnergy_{\mathrm{inc}}(\feps,z\,| z_n) = \frac{1}{2z}\feps:\C:\feps + \triangle t\, \dissipot \left( \frac{z-z_n}{\triangle t\,m(z_n)} \right).
\end{equation}
In contrast to the fully implicit potential~\eqref{eq:theory_frozen_metric_fully_implicit_potential_explicit}, the metric $m$ is evaluated at the known state $z_n$. Thus, the argument of the dissipation potential in Eq.~\eqref{eq:theory_frozen_metric_incremental_potential_explicit} is affine in the current state $z$. Hence, the representation~\eqref{eq:theory_frozen_metric_incremental_potential_explicit} reveals the frozen-metric discretization to be jointly convex in the strain $\feps$ and the reciprocal degradation state $z$ for every positive time step size. Indeed, the transformed free energy~\eqref{eq:theory_hidden_convexity_transformed_energy} is jointly convex, whereas the dissipation contribution in Eq.~\eqref{eq:theory_frozen_metric_incremental_potential_explicit} is convex in the variable $z$ because its argument is affine, see Appendix~\ref{apx:frozen_metric_convexity}.

The effective domain of the original dissipation potential~\eqref{eq:theory_scalar_model_dissipation} implies that the incremental potential~\eqref{eq:theory_frozen_metric_incremental_potential_explicit} is finite precisely for the states satisfying the inequalities
\begin{equation}\label{eq:theory_frozen_metric_admissibility_preliminary}
	0 \leq \frac{z - z_n}{\triangle t \, m(z_n)} < \ddotmax.
\end{equation}
Equivalently, the admissible interval for the current reciprocal degradation state is characterized by the bounds
\begin{equation}\label{eq:theory_frozen_metric_admissibility}
	z_n \leq z < z_n + m(z_n) \, \triangle t \, \ddotmax.
\end{equation}
We summarize the constitutive and mathematical properties of the frozen-metric discretization~\eqref{eq:theory_frozen_metric_incremental_potential} in the following. The arguments for the convexity, the update, the Newton iteration and the tangent are provided in Appendix~\ref{apx:frozen_metric_properties}.
\begin{enumerate}
	\item The Cauchy stress associated with the incremental potential~\eqref{eq:theory_frozen_metric_incremental_potential_explicit} computes as
	\begin{equation}\label{eq:theory_frozen_metric_stress}
		\fsigma := \frac{\partial\freeEnergy_{\mathrm{inc}}}{\partial\feps}(\feps,z\,| z_n) = \frac{1}{z} \, \C:\feps.
	\end{equation}
	As the update minimizes the incremental potential in its second argument, this partial derivative coincides with the derivative of the condensed incremental potential.
	\item The frozen-metric update preserves both damage irreversibility and the prescribed bound on the rate of the original damage variable. By the admissible interval~\eqref{eq:theory_frozen_metric_admissibility}, every state with a finite incremental potential satisfies the inequality $z\geq z_n$. As the inverse function $g^{-1}$ is strictly monotonically increasing, the corresponding damage update is irreversible,
	\begin{equation}\label{eq:theory_frozen_metric_irreversibility}
		d_{n+1}=g^{-1}(z)\geq g^{-1}(z_n)=d_n.
	\end{equation}
	Moreover, the function $g^{-1}$ lies below its tangent at the previous state, whose slope is the reciprocal metric~\eqref{eq:theory_hidden_convexity_metric}, so that the admissible interval~\eqref{eq:theory_frozen_metric_admissibility} implies
	\begin{equation}\label{eq:theory_frozen_metric_physical_rate_bound}
		0\leq d_{n+1}-d_n\leq\frac{z-z_n}{m(z_n)}<\triangle t\,\ddotmax.
	\end{equation}
	In particular, the reduced discrete Clausius--Duhem inequality~\cite{Schneider2026_Redistribution}
	\begin{equation}\label{eq:theory_frozen_metric_Clausius_Duhem}
		0 \leq -\frac{\partial\widehat{\freeEnergy}}{\partial z}(\feps,z)[z-z_n] = \frac{\freeEnergy_0(\feps)}{z^2}(z-z_n)
	\end{equation}
	is satisfied. By the rate bound~\eqref{eq:theory_frozen_metric_physical_rate_bound}, the frozen metric charges the dissipation for a damage increment which exceeds the actual one, and it evaluates the driving force with the metric at the previous state, which is nondecreasing by the convexity assumption~\eqref{eq:theory_scalar_model_regularities}. Both effects slow down the damage evolution, so that the damage of the frozen-metric update lags behind the scalar evolution~\eqref{eq:theory_scalar_model_evolution} and the stress on the softening branch is overestimated, see Sec.~\ref{sec:numerics_material_point}.
	\item For a prescribed strain $\feps$, the updated reciprocal degradation state is determined by the minimization problem
	\begin{equation}\label{eq:theory_frozen_metric_update}
		z := \argmin \left\{ \freeEnergy_{\mathrm{inc}}(\feps, \zeta \, | z_n) \, \middle| \, z_n \leq \zeta < z_n + m(z_n) \triangle t \, \ddotmax \right\}.
	\end{equation}
	The update remains inactive, i.e., the state stays at the previous state $z_n$, precisely if the condition
	\begin{equation}\label{eq:theory_frozen_metric_inactive_condition}
		m(z_n) \, \frac{\freeEnergy_0(\feps)}{z_n^2} \leq	r + \lambda
	\end{equation}
	holds.
	Otherwise, the updated state is the unique solution of the scalar equation
	\begin{equation}\label{eq:theory_frozen_metric_scalar_equation}
		\dissipot'\left( \frac{z-z_n}{\triangle t\,m(z_n)} \right) = m(z_n) \, \frac{\freeEnergy_0(\feps)}{z^2}.
	\end{equation}
	The residual associated with Eq.~\eqref{eq:theory_frozen_metric_scalar_equation} is strictly monotonically increasing on the admissible interval~\eqref{eq:theory_frozen_metric_admissibility}, as its derivative
	\begin{equation}\label{eq:theory_frozen_metric_residual_derivative}
		\frac{\dissipot''(s)}{\triangle t\,m(z_n)} + \frac{2\,m(z_n)\,\freeEnergy_0(\feps)}{z^3} > 0, \qquad s := \frac{z-z_n}{\triangle t\,m(z_n)},
	\end{equation}
	is positive by Eq.~\eqref{eq:theory_scalar_model_dissipation_derivatives}, and it diverges at its upper endpoint, see Eq.~\eqref{eq:theory_scalar_model_dissipation_derivatives}, whereas it is negative at the previous state $z_n$ precisely if the inactive condition~\eqref{eq:theory_frozen_metric_inactive_condition} is violated. Thus, the active update exists and is unique. For the dissipation potential~\eqref{eq:theory_scalar_model_dissipation}, the scalar equation~\eqref{eq:theory_frozen_metric_scalar_equation} is equivalent in the admissible interval to the vanishing of the pole-free residual
	\begin{equation}\label{eq:theory_frozen_metric_pole_free_residual}
		M(z) := \lambda + \left( 1-\frac{z-z_n}{\alpha_n} \right) \left( r-\frac{A_n}{z^2} \right),
	\end{equation}
	where we introduce the length $\alpha_n$ of the admissible interval~\eqref{eq:theory_frozen_metric_admissibility} and the undamaged energy $A_n$ scaled by the metric at the previous state,
	\begin{equation}\label{eq:theory_frozen_metric_scalar_equation_abbreviations}
		\alpha_n:=\triangle t\,m(z_n) \, \ddotmax \quad\text{and}\quad	A_n:=m(z_n) \, \freeEnergy_0(\feps).
	\end{equation}
	For the residual~\eqref{eq:theory_frozen_metric_pole_free_residual}, the undamped Newton iteration initialized by the previous state $z_n$ remains in the admissible interval and converges monotonically from below to the unique active solution $z_\ast$. The arguments are provided in Appendix~\ref{apx:frozen_metric_Newton}.
	\item On the active branch, the consistent algorithmic material tangent takes the form
	\begin{equation}\label{eq:theory_frozen_metric_material_tangent}
		\C_{\mathrm{alg}} =	\frac{1}{z} \left[ \C -	\frac{\triangle t\,\ddotmax \gamma}{\lambda z^3 + \triangle t\,\ddotmax \, (\feps:\C:\feps) \, \gamma	} (\C:\feps)\otimes(\C:\feps)
		\right],
	\end{equation}
	with the factor
	\begin{equation}\label{eq:theory_frozen_metric_tangent_factor}
		\gamma := [m(z_n)]^2 \left(1-\frac{z-z_n}{\triangle t\,m(z_n) \, \ddotmax}\right)^2
	\end{equation}
	which depends on the current state $z$ as well, and where this state is the unique solution of the incremental minimization problem~\eqref{eq:theory_frozen_metric_update}. On the inactive branch, the reciprocal degradation variable remains fixed at the previous state, and the material tangent reduces to the damaged elastic tangent
	\begin{equation}\label{eq:theory_frozen_metric_elastic_tangent}
		\C_{\mathrm{alg}} =	\frac{1}{z_n}\C.
	\end{equation}
	The tangent is symmetric and uniformly positive definite with respect to the current strain, see Appendix~\ref{apx:frozen_metric_tangent}, the uniformity referring to the strain at a fixed previous state and time step size. More precisely, the estimate
	\begin{equation}\label{eq:theory_frozen_metric_uniform_tangent_bound}
		\fxi:\C_{\mathrm{alg}}:\fxi \geq \frac{\lambda}{\lambda z_n+2\alpha_n(r+\lambda)}\,\fxi:\C:\fxi, \qquad \fxi\in\Sym{3},
	\end{equation} holds on both the active and inactive branches. Consequently, the incremental potential condensed over the reciprocal degradation state is strongly convex in the strain. The factor in the bound~\eqref{eq:theory_frozen_metric_uniform_tangent_bound} decreases as the previous state or the time step size grows, as the metric is nondecreasing by the convexity assumption~\eqref{eq:theory_scalar_model_regularities}, and it stays bounded away from zero only on bounded sets of both. By the rate bound~\eqref{eq:theory_frozen_metric_physical_rate_bound}, the estimate
	\begin{equation}\label{eq:theory_frozen_metric_state_bound}
		z_n\leq g(d_0+T\ddotmax)
	\end{equation}
	holds at every material point, finite by the condition~\eqref{eq:theory_scalar_model_attainable_damage_range}. The factor in the bound~\eqref{eq:theory_frozen_metric_uniform_tangent_bound} thus stays bounded away from zero uniformly over the domain, whereas the tangent~\eqref{eq:theory_frozen_metric_material_tangent} is bounded above by the undamaged elasticity tensor $\C$, so that the incremental boundary value problem is well-posed in the sense of Hadamard, as for Entropic Damage Transport~\cite{Lauff2026_OTRModel}.
\end{enumerate}
A possible implementation is sketched in Alg.~\ref{alg:theory_frozen_metric_update}. As the Newton iteration for the pole-free residual~\eqref{eq:theory_frozen_metric_pole_free_residual} converges monotonically by the invariant~\eqref{eq:apx_frozen_metric_Newton_invariant}, it needs neither damping nor bisection safeguards, and in the elastic domain the algorithm performs no iteration at all. The residual criterion is scaled by the characteristic magnitude of the terms in the pole-free residual, whereas the second criterion controls the relative Newton increment.

\begin{algorithm}[ht]\caption{Frozen-metric material-point update for the scalar Entropic Damage-Transport model}\label{alg:theory_frozen_metric_update}
\begin{algorithmic}[1]
	\Require $\feps$, $z_n$, $\triangle t$, $\C$, $r$, $\lambda$, $\ddotmax$, $m$, $g^{-1}$, $\mathrm{tol}$, $\mathrm{maxit}$
	\Ensure $z$, $d$, $\fsigma$, and $\C_{\mathrm{alg}}$ if required
	\State $\freeEnergy_0\leftarrow\frac{1}{2}\feps:\C:\feps$, $m_n\leftarrow m(z_n)$ \Comment{Eqs.~\eqref{eq:theory_scalar_model_undamaged_energy} and~\eqref{eq:theory_hidden_convexity_metric}}
	\State $\alpha_n\leftarrow\triangle t\,m_n\ddotmax$, $A_n\leftarrow m_n\freeEnergy_0$ \Comment{Abbreviations~\eqref{eq:theory_frozen_metric_scalar_equation_abbreviations}}
	\If{$A_n/z_n^2\leq r+\lambda$} \Comment{Inactive update, Eq.~\eqref{eq:theory_frozen_metric_inactive_condition}}
		\State $z\leftarrow z_n$, $d\leftarrow g^{-1}(z)$
		\State $\fsigma\leftarrow z^{-1}\C:\feps$, $\C_{\mathrm{alg}}\leftarrow z^{-1}\C$ \Comment{Eqs.~\eqref{eq:theory_frozen_metric_stress} and~\eqref{eq:theory_frozen_metric_elastic_tangent}}
		\State \Return $z,d,\fsigma,\C_{\mathrm{alg}}$
	\EndIf
	\State $z\leftarrow z_n$ \Comment{Newton initialization}
	\For{$\mathrm{iteration}=1,2,\ldots,\mathrm{maxit}$}
		\State $M_z\leftarrow M(z)$, $M_z'\leftarrow M'(z)$ \Comment{Residual~\eqref{eq:theory_frozen_metric_pole_free_residual}}
		\State $M_{\mathrm{scale}}\leftarrow\lambda+\left(1-\dfrac{z-z_n}{\alpha_n}\right)\left(r+\dfrac{A_n}{z^2}\right)$ \Comment{Magnitude of the terms of the residual}
		\If{$|M_z|\leq\mathrm{tol}\,M_{\mathrm{scale}}$} \Comment{Residual criterion}
			\State \textbf{break}
		\EndIf
		\State $\triangle z\leftarrow-M_z/M_z'$ \Comment{Undamped Newton increment}
		\State $z\leftarrow z+\triangle z$
		\If{$|\triangle z|\leq\mathrm{tol}\,$$z$} \Comment{Increment criterion, $z\geq1$}
			\State \textbf{break}
		\EndIf
	\EndFor
	\State $d\leftarrow g^{-1}(z)$, $\fsigma\leftarrow z^{-1}\C:\feps$ \Comment{Optional: compute $\C_{\mathrm{alg}}$ from Eq.~\eqref{eq:theory_frozen_metric_material_tangent}}
	\State \Return $z,d,\fsigma,\C_{\mathrm{alg}}$
\end{algorithmic}
\end{algorithm}

Each material point carries a single scalar state variable, and both the stress~\eqref{eq:theory_frozen_metric_stress} and the tangent~\eqref{eq:theory_frozen_metric_material_tangent} are available in closed form, so that the frozen-metric model is incorporated directly into conventional Newton-based finite-element and computational homogenization frameworks.

The frozen-metric update~\eqref{eq:theory_frozen_metric_update} is a time discretization of the scalar evolution~\eqref{eq:theory_scalar_model_evolution} in its own right. At material-point level, it converges to the scalar evolution at first order in the time step size, see Appendix~\ref{apx:frozen_metric_convergence}.

\section{\texorpdfstring{Computational investigations}{Computational investigations}}
\label{sec:numerics}

\subsection{Setup}
\label{sec:numerics_setup}

In this section, we assess the frozen-metric discretization of the scalar Entropic Damage-Transport model (sEDT) introduced in Sec.~\ref{sec:theory_frozen_metric} against the measure-valued Entropic Damage Transport (EDT), from the material point up to an industrial component. We keep the name of Entropic Damage Transport for the scalar model, as its evolution law~\eqref{eq:theory_scalar_model_evolution} with the entropic regularization parameter $\lambda$ is precisely the law by which EDT moves its weights between neighboring damage states, see Eq.~\eqref{eq:apx_continuous_limit_volume_fractions}. Except for the component of Sec.~\ref{sec:numerics_bracket}, which carries the titanium alloy of the data set~\cite{Whalen2021}, all computations use the material data of the fiber-reinforced polyamide listed in Tab.~\ref{tab:numerics_setup_materials}. We employ the exponential degradation~\eqref{eq:theory_scalar_model_exponential_degradation}, truncated at the residual degradation $f_{\mathrm{res}}$ listed in Tab.~\ref{tab:numerics_setup_materials}, equivalent to one percent of the undamaged stiffness, i.e., at the maximum damage $\dmax=-\log f_{\mathrm{res}}\approx4.61$. The damage-rate limitation $\ddotmax$ and the duration of the load paths are chosen such that the attainable-damage condition~\eqref{eq:theory_scalar_model_attainable_damage_range} holds for this maximum damage. Hence, the truncation never becomes active, and the assumptions of Sec.~\ref{sec:theory} hold along all load paths, compare the smallest degradation of $0.0127$ attained in the component, see Tab.~\ref{tab:bracket_ladder}.

\begin{table}[ht]
	\centering
	\caption{Material parameters used for the computational investigations. The elastic constants of the matrix and of the inclusion are taken from Lauff et al.~\cite{Lauff2026_OTRModel}, the damage parameters of the matrix from Schneider~\cite{Schneider2026_Redistribution}, the titanium alloy of the component of Sec.~\ref{sec:numerics_bracket} from the data set~\cite{Whalen2021}. Only the matrix degrades, whereas the inclusions remain linearly elastic. Where two values of a damage parameter are given, the first one refers to the material-point investigations of Sec.~\ref{sec:numerics_material_point} and to the notched specimen of Sec.~\ref{sec:numerics_notched_specimen} and the second one to the fiber-reinforced cell of Sec.~\ref{sec:numerics_unit_cell}.}
	\label{tab:numerics_setup_materials}
	\small
	\setlength{\tabcolsep}{4.5pt}
	\begin{tabular}{llcccc}
		\toprule
		& & units & matrix & inclusion & component\\
		& & & (polyamide) & (E-glass) & (Ti-6Al-4V)\\
		\midrule
		Young's modulus & $E$ & \unit{\giga\pascal} & $3$ & $72$ & $113.8$\\
		Poisson's ratio & $\nu$ & -- & $0.4$ & $0.22$ & $0.342$\\
		\midrule
		damage-initiation threshold & $r$ & \unit{\mega\pascal} & $0.1$ / $0.45$ & -- & $3.2584$\\
		entropic regularization parameter & $\lambda$ & \unit{\mega\pascal} & $0.01$ / $0.05$ & -- & $0.32584$\\
		damage-rate limitation & $\ddotmax$ & \unit{\per\second} & $1$ / $4$ & -- & $4$\\
		degradation tolerance & $f_{\mathrm{res}}$ & -- & $10^{-2}$ & -- & $10^{-2}$\\
		\bottomrule
	\end{tabular}
\end{table}

The frozen-metric update of Alg.~\ref{alg:theory_frozen_metric_update} uses a tolerance of $10^{-10}$ and the iteration limit $\mathrm{maxit}$ of $50$. EDT solves its material-point problem to the same relative tolerance. At material-point level and for the notched specimen, we employ the redistribution strategy~\cite{Schneider2026_Redistribution} onto ten damage states. On the unit cell, we compare with the original EDT without redistribution, whose number of damage states grows with the number of time steps, as the redistribution introduces an additional approximation of the distribution, which would blur the comparison of the two time discretizations. For EDT, we use the Newton-bisection material-point solver~\cite{Schneider2026FastOTR} throughout, which solves the same subproblem as the projected-gradient scheme~\cite{Lauff2026_OTRModel} but is faster. We solve the balance of linear momentum on the unit cell of Sec.~\ref{sec:numerics_unit_cell} with an in-house FFT-based computational micromechanics code based on the Lippmann--Schwinger equation~\cite{MoulinecSuquet1994, MoulinecSuquet1998}. The problem is discretized by finite differences on a staggered grid~\cite{SchneiderOspaldKabel2016}. We use the Newton-CG solver~\cite{Gelebart2013,Kabel2014,Wicht2020} and the convergence criterion of Schneider~\cite{SchneiderReview2021} with a tolerance of $10^{-5}$. The solver requires the consistent algorithmic material tangent in each Newton step, which we derive in Appendix~\ref{apx:frozen_metric_tangent}.

The number of equidistant load steps is denoted by $n_{\triangle t}$. Accuracy is measured using the normalized root mean square error (NRMSE) of a scalar discrete history $\chi$, in all studies below an axial stress component, sampled at the load steps $t^n$ against a reference \mbox{history $\chi_{\mathrm{ref}}$},
\begin{equation}
\label{eq:numerics_setup_error}
	\mathrm{NRMSE}(\chi)
	=
	\displaystyle
	\frac{1}{\displaystyle\max_{t\in\mathcal{T}_{\mathrm{ref}}}|\chi_{\mathrm{ref}}(t)|}
	\sqrt{
		\frac{1}{n_{\triangle t}+1}
		\sum_{n=0}^{n_{\triangle t}}
		\left(
			\chi(t^n)-\chi_{\mathrm{ref}}(t^n)
		\right)^2
	},
\end{equation}
which we report in percent, where $\mathcal{T}_{\mathrm{ref}}$ denotes the instants of the reference history. The normalization by the peak of the reference measures the deviation relative to the load-carrying capacity, which remains meaningful on the softening branch, where the reference stress itself decreases. The reference history is chosen individually for each of the studies below. We run the material-point tests of Sec.~\ref{sec:numerics_material_point} and the tension tests of the notched specimen of Sec.~\ref{sec:numerics_notched_specimen} on an Apple M2 with four performance cores and \qty{24}{\giga\byte} of unified memory. The unit-cell simulations of Sec.~\ref{sec:numerics_unit_cell} and the component of Sec.~\ref{sec:numerics_bracket} run on a dual-socket workstation carrying two AMD EPYC 9354 processors with a total of $64$ physical cores and \qty{1.1}{\tera\byte} of memory. Thus, within any section, the runtimes are ensured to be comparable. On the unit cell, we measure the runtimes on a single thread and average over five runs.

\subsection{Investigations at material-point level}
\label{sec:numerics_material_point}

We first investigate the accuracy of the frozen-metric time discretization at material-point level. We implement both models as user materials (\texttt{UMAT}) in Abaqus and load a single eight-node hexahedral element with reduced integration (C3D8R) in uniaxial tension under small deformations. We prescribe the axial strain $\varepsilon_{11}$, which increases linearly from zero to \qty{3}{\percent} within \qty{3}{\second}, and keep the lateral faces traction-free.

\begin{figure}[b!ht]
	\centering
	\captionsetup[subfigure]{font=normalsize}
	\includegraphics{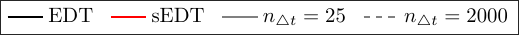}\\[0.8em]
	\begin{subfigure}[t]{0.48\textwidth}
		\centering
		\includegraphics[width=\linewidth]{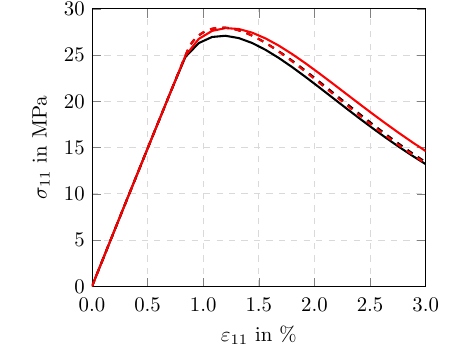}
		\caption{stress--strain response}
		\label{fig:numerics_material_point_response}
	\end{subfigure}
	\hspace{1em}
	\begin{subfigure}[t]{0.48\textwidth}
		\centering
		\includegraphics[width=\linewidth]{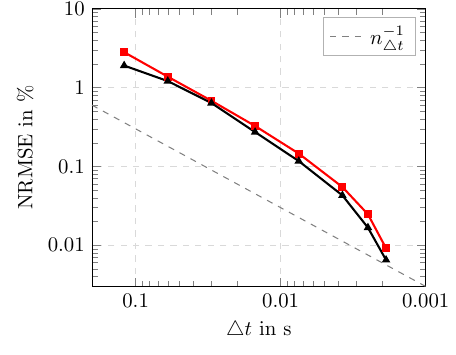}
		\caption{stress error under temporal refinement}
		\label{fig:numerics_material_point_error}
	\end{subfigure}
	\caption{Uniaxial tension at a material point. (a) Stress--strain response of EDT and sEDT. (b) Error~\eqref{eq:numerics_setup_error} of the axial stress over the time step size, each model against its own response for $n_{\triangle t}=2000$ load steps.}
	
	\label{fig:numerics_material_point}
\end{figure}

We vary the number $n_{\triangle t}$ of load steps from $25$ to $2000$. Fig.~\ref{fig:numerics_material_point_response} shows the responses for the coarsest and the finest discretization. Both models predict the same damage-initiation strain of about \qty{0.85}{\percent}. For $n_{\triangle t}=2000$ load steps, the responses of both models nearly coincide, as both converge to the same scalar evolution, see Appendix~\ref{apx:convergence}. Relative to these responses, EDT underestimates the peak stress by approximately \qty{3}{\percent} for $n_{\triangle t}=25$ load steps, but approaches the finely resolved response again during softening and nearly coincides with it at the end of the loading. In contrast, sEDT reproduces the peak stress at an axial strain of about \qty{1.2}{\percent} to within \qty{0.2}{\percent}, but overestimates the stress along the softening branch, by approximately \qty{10}{\percent} of the local stress, i.e., by \qty{4.7}{\percent} of the peak stress, at the final strain of \qty{3}{\percent}. For EDT, the stress error introduced around the peak decreases again along the softening branch, whereas for sEDT, the error of each step accumulates there, as the frozen metric lets the damage lag behind the scalar evolution, see Sec.~\ref{sec:theory_frozen_metric}.

Fig.~\ref{fig:numerics_material_point_error} shows the error~\eqref{eq:numerics_setup_error} of each model against its own response for $n_{\triangle t}=2000$ load steps. For both models, the error remains below \qty{3}{\percent} even for $n_{\triangle t}=25$ load steps and drops to about \qty{0.3}{\percent} for $200$ load steps. The errors of both models are of similar magnitude and decrease at about the same rate, approximately first order in the time step size. This rate agrees with the estimate~\eqref{eq:theory_frozen_metric_convergence_estimate} for sEDT. For EDT, it exceeds the half order guaranteed by the upper bound~\eqref{eq:theory_scalar_model_continuous_limit_estimate}.

\subsection{Computational performance at unit-cell level}
\label{sec:numerics_unit_cell}

\begin{figure}[ht]
	\centering
	\captionsetup[subfigure]{font=normalsize}
	\begin{subfigure}[b]{0.44\textwidth}
		\centering
		\includegraphics{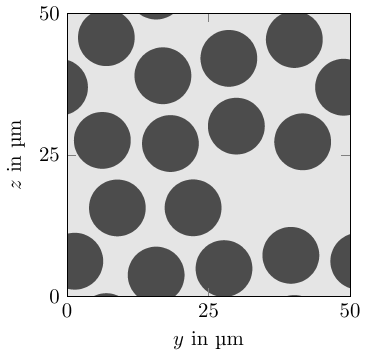}
		\caption{microstructure}
		\label{fig:numerics_unit_cell_setup_microstructure}
	\end{subfigure}
	\hfill
	\begin{subfigure}[b]{0.50\textwidth}
		\centering
		\includegraphics{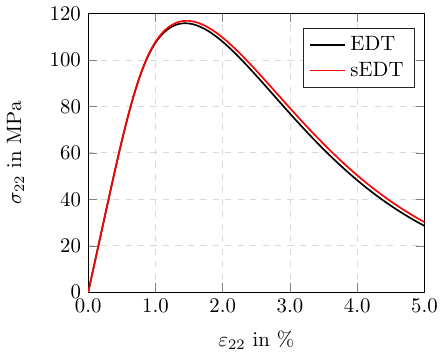}
		\caption{stress--strain response}
		\label{fig:numerics_unit_cell_setup_response}
	\end{subfigure}
	\caption[Microstructure and stress--strain response of the unidirectionally fiber-reinforced unit cell]{Microstructure and stress--strain response of the unidirectionally fiber-reinforced unit cell at the resolution of $1600^2$ pixels, with the polyamide matrix in light and the E-glass fibers in dark.}
	
	\label{fig:numerics_unit_cell_setup}
\end{figure}

We consider the unidirectionally fiber-reinforced composite shown in Fig.~\ref{fig:numerics_unit_cell_setup_microstructure}. The fiber arrangement was generated by the mechanical contraction method~\cite{WilliamsPhilipse2003} at a minimum fiber distance of \qty{2}{\micro\metre}. The two-dimensional cell has the edge length of \qty{50}{\micro\metre} and contains $15$ fibers of diameter of \qty{10}{\micro\metre}, corresponding to a fiber volume fraction of \qty{47.1}{\percent}. Only the matrix degrades, whereas the fibers remain linearly elastic, see Tab.~\ref{tab:numerics_setup_materials}. The cell is loaded by the macroscopic normal strain $\varepsilon_{22}$ up to \qty{5}{\percent} within \qty{1}{\second}, with all other macroscopic strain components held at zero. We employ periodic boundary conditions. Both formulations initiate damage once the driving force exceeds the total threshold $r+\lambda$ of \qty{0.5}{\mega\pascal}, see Eq.~\eqref{eq:theory_scalar_model_elastic_domain}, which we split into the damage-initiation threshold $r$ and the entropic parameter $\lambda$. Tab.~\ref{tab:numerics_setup_materials} lists the values of all parameters of the damage model.

\begin{figure}[ht]
	\centering
	\includegraphics{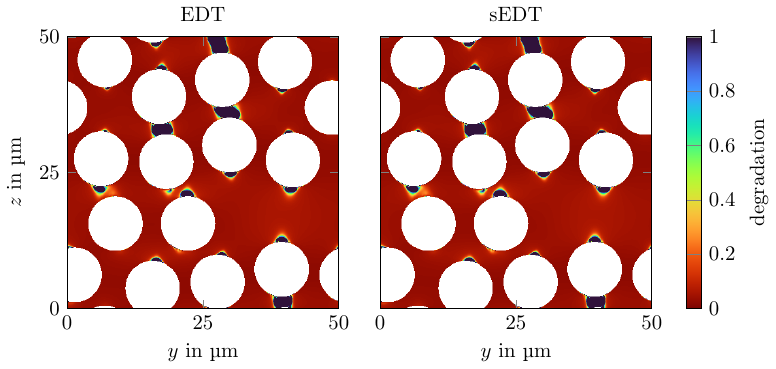}
	\caption{The local degradation, $\bar f$ for EDT and $f$ for sEDT, of the two formulations at the end of the load path, computed using $N_i=400$ pixels per edge length and a time step size of $\triangle t =$ \qty{1.5625}{\milli\second}. The fibers are left white as they are modeled linearly elastic.}
	\label{fig:numerics_unit_cell_fields}
\end{figure}

We refine the cell in two respects separately. In space we use $16$ values of the pixel count $N_i$ per axis, logarithmically equidistant between $50$ and $1600$ and rounded to numbers whose prime factors are at most five, which keeps the fast Fourier transform at its optimal complexity, and we fix the time step size at \qty{10}{\milli\second}. In time we fix the spatial resolution $N_i$ at $100$ and let the number of load steps $n_{\triangle t}$ run over $110$ logarithmically spaced values between $5$ and $640$. Accuracy is measured by the normalized root mean square error~\eqref{eq:numerics_setup_error} of the axial stress $\sigma_{22}$, each formulation against its own finest response. At the resolution of $1600^2$ pixels, the two formulations separate by at most \qty{2.02}{\percent} of the peak stress, see Fig.~\ref{fig:numerics_unit_cell_setup_response}. Under temporal refinement at the pixel count $N_i=100$ per axis, the distance between the two responses, measured by the error~\eqref{eq:numerics_setup_error} of one against the other, decreases at first order, from \qty{24}{\percent} for a time step size of \qty{0.2}{\second} to \qty{0.19}{\percent} for \qty{1.25}{\milli\second}. At equal resolution and time step size, the degradation fields of both formulations differ by \num{0.0012} in the mean over the matrix, see Fig.~\ref{fig:numerics_unit_cell_fields}.

\begin{figure}[htp]
	\centering
	\captionsetup[subfigure]{font=normalsize}
	\includegraphics{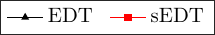}\\[0.8em]
	\begin{subfigure}[t]{0.48\textwidth}
		\centering
		\includegraphics[width=\textwidth]{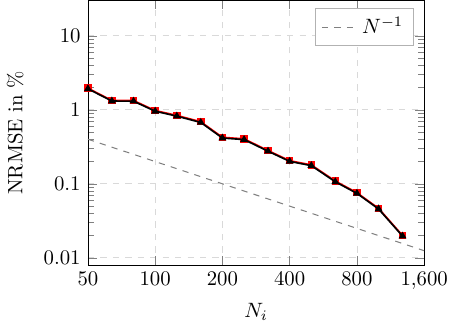}
		\caption{spatial refinement}
		\label{fig:numerics_unit_cell_accuracy_spatial}
	\end{subfigure}
	\hfill
	\begin{subfigure}[t]{0.48\textwidth}
		\centering
		\includegraphics[width=\textwidth]{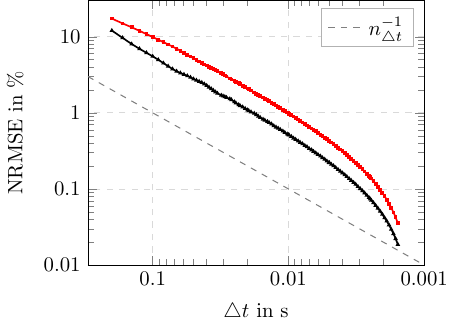}
		\caption{temporal refinement}
		\label{fig:numerics_unit_cell_accuracy_temporal}
	\end{subfigure}
	\caption{Accuracy for the unidirectionally reinforced cell under the two refinements. (a) Spatial refinement at the time step size of \qty{10}{\milli\second}. (b) Temporal refinement at the fixed resolution $N_i=100$.}
	
	\label{fig:numerics_unit_cell_accuracy}
\end{figure}

\begin{figure}[htp]
	\centering
	\captionsetup[subfigure]{font=normalsize}
	\includegraphics{plots/f1_resolution/cell-legend.pdf}\\[0.8em]
	\begin{subfigure}[t]{0.48\textwidth}
		\centering
		\includegraphics[width=\textwidth]{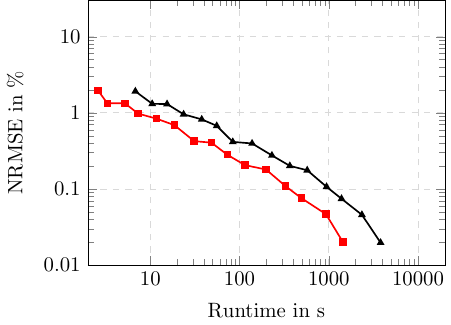}
		\caption{spatial refinement}
		\label{fig:numerics_unit_cell_cost_spatial}
	\end{subfigure}
	\hfill
	\begin{subfigure}[t]{0.48\textwidth}
		\centering
		\includegraphics[width=\textwidth]{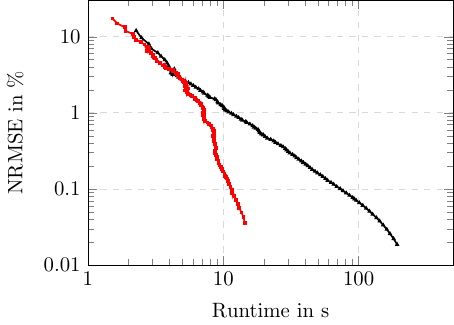}
		\caption{temporal refinement}
		\label{fig:numerics_unit_cell_cost_temporal}
	\end{subfigure}
	\caption{Accuracy against runtime for the unidirectionally reinforced cell, one point per entry of the two ladders, with the error of Fig.~\ref{fig:numerics_unit_cell_accuracy} over the total runtime. (a) Spatial refinement, ending at $N_i=1280$. (b) Temporal refinement, ending at $n_{\triangle t}=640$.}
	\label{fig:numerics_unit_cell_cost}
\end{figure}

Fig.~\ref{fig:numerics_unit_cell_accuracy} compares the accuracy of the two formulations under the two refinements. Under spatial refinement, the errors of sEDT, each measured against its own finest response, decrease in the same way as those of EDT, see Fig.~\ref{fig:numerics_unit_cell_accuracy_spatial}. At the fixed time step size, the two spatial limits differ by the distance of the two time discretizations, which vanishes only under temporal refinement. The two errors agree to within a relative deviation of \qty{2.5}{\percent} at every resolution, and both fall at first order. At least $100$ pixels per axis are needed to bring the spatial error below \qty{1}{\percent}.

Under temporal refinement, both formulations converge at first order, see Fig.~\ref{fig:numerics_unit_cell_accuracy_temporal}. Here the two curves do not coincide but run parallel, sEDT carrying a factor of $1.9$ more error at the same time step size, between $1.4$ and $2.0$ over the ladder. Reaching an error of \qty{1}{\percent} therefore takes $55$ load steps for EDT and $99$ for sEDT.

The iteration counts of the homogenization solver do not show that difference. Bounded under grid refinement for both formulations, they fall between the resolutions of $N_i=50$ and $N_i=1600$ pixels per axis from $28.5$ to $24.4$ for sEDT and from $25.1$ to $22.5$ for EDT, so that sEDT inherits the resolution-independent conditioning of EDT at about a tenth more iterations per load step.

The two formulations differ in the cost of the material law, as an evaluation of EDT updates all damage states of the distribution, whereas an evaluation of sEDT updates a single scalar. Plotted against the total runtime, see Fig.~\ref{fig:numerics_unit_cell_cost}, sEDT reaches the spatial error of \qty{0.1}{\percent} at a factor of $2.8$ less runtime, and under temporal refinement its advantage grows with the accuracy demanded, from parity at \qty{3}{\percent} to a factor of $6.4$ at \qty{0.1}{\percent}. EDT also requires significantly more memory, as it carries the damage measure on a grid of damage states whose number grows as the time step size decreases. At the finest resolution $N_i=1600$, its internal variables occupy $117$ times the memory of the single scalar per pixel of sEDT, which raises the total memory by a factor of $1.9$. These factors refer to the original EDT. The redistribution~\cite{Schneider2026_Redistribution} onto ten damage states bounds the internal variables of EDT independently of the time step size, still at least ten times the single scalar of sEDT, and reduces the cost per material point, at the price of an additional approximation of the distribution.

\subsection{Single edge notched tension specimen}
\label{sec:numerics_notched_specimen}

\begin{figure}[ht]
	\centering
	\includegraphics[width=\textwidth]{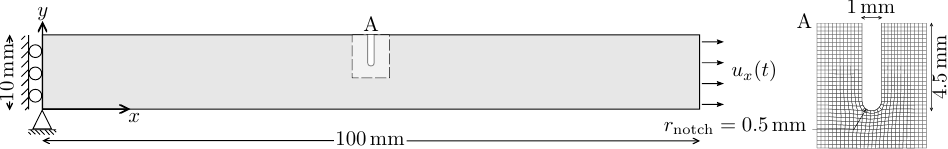}
	\caption{Geometry of the single edge notched tension (SENT) specimen with boundary conditions and the prescribed displacement $u_x(t)$. The detail~A around the notch is shown enlarged on the right, with the mesh of element size $\ell_{\mathrm{elem}} = \qty{0.2}{\milli\metre}$.}
	\label{fig:sent_specimen}
\end{figure}

As a simple component, we consider the single edge notched tension (SENT) specimen of Fig.~\ref{fig:sent_specimen} with the matrix data of Tab.~\ref{tab:numerics_setup_materials}. The left end is fixed in $x$-direction, its lower corner also in $y$-direction, and the displacement $u_x$ of the right end increases linearly to \qty{3}{\milli\metre} within \qty{3}{\second} in $n_{\triangle t}=50$ load steps. The front and back faces are fixed in $z$-direction, so that the specimen deforms in plane strain. We use the \texttt{UMAT} implementations of Sec.~\ref{sec:numerics_material_point} under small deformations, as the strain of the specimen remains below \qty{3}{\percent}, and C3D8R elements, two of them across the thickness of \qty{1}{\milli\metre}. We consider five meshes with in-plane element sizes $\ell_{\mathrm{elem}}$ from \qty{0.5}{\milli\metre} to \qty{0.1}{\milli\metre}.

\begin{figure}[ht]
	\centering
	\captionsetup[subfigure]{font=normalsize}
	\includegraphics{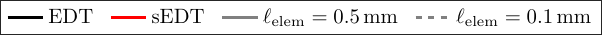}\\[0.8em]
	\begin{subfigure}[t]{0.48\textwidth}
		\centering
		\includegraphics[width=\textwidth]{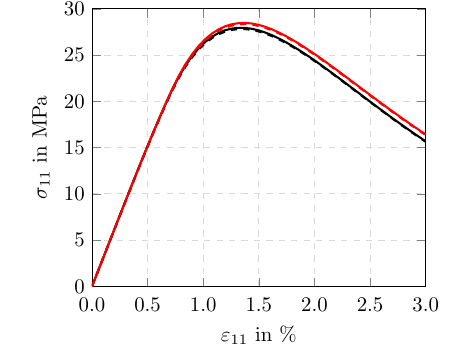}
		\caption{stress--strain plot}
		\label{fig:sent_plot_stress}
	\end{subfigure}
	\hspace{1em}
	\begin{subfigure}[t]{0.48\textwidth}
		\centering
		\includegraphics[width=\textwidth]{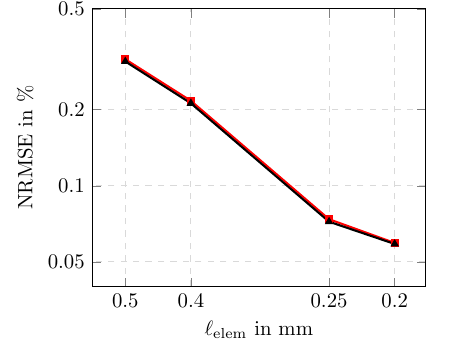}
		\caption{error under spatial refinement}
		\label{fig:sent_plot_nrmse_stress}
	\end{subfigure}
	\caption[Stress--strain response of the SENT specimen and its error under spatial refinement]{Stress--strain response of the SENT specimen under tension on two meshes and error~\eqref{eq:numerics_setup_error} of the stress under spatial refinement, each model against its own response on the finest mesh.}
	
	\label{fig:sent_plot_stress_error}
\end{figure}

Fig.~\ref{fig:sent_plot_stress} shows the stress--strain response for different meshes, where the stress $\sigma_{11}$ is the reaction force divided by the gross cross section of \qty{10}{\milli\metre\squared}, and the strain $\varepsilon_{11}$ is the prescribed displacement divided by the specimen length. Fig.~\ref{fig:sent_plot_nrmse_stress} shows the error~\eqref{eq:numerics_setup_error} of the stress under spatial refinement with respect to the finest discretization with element size $\ell_{\mathrm{elem}} = \qty{0.1}{\milli\metre}$.
Under mesh refinement, the error decreases from about \qty{0.3}{\percent} on the coarsest mesh with element size \qty{0.5}{\milli\metre} to \qty{0.06}{\percent} on the \qty{0.2}{\milli\metre} mesh, for both models alike. In contrast, the stress of sEDT exceeds that of EDT by about \qty{2}{\percent} at the peak and by about \qty{5}{\percent} at the end of the load path. This difference conforms with the temporal discretization error at material-point level for a comparable step size, where sEDT overestimates the softening stress and EDT underestimates the peak, see Fig.~\ref{fig:numerics_material_point}.

\begin{figure}[ht]
	\centering
	\includegraphics[width=\textwidth]{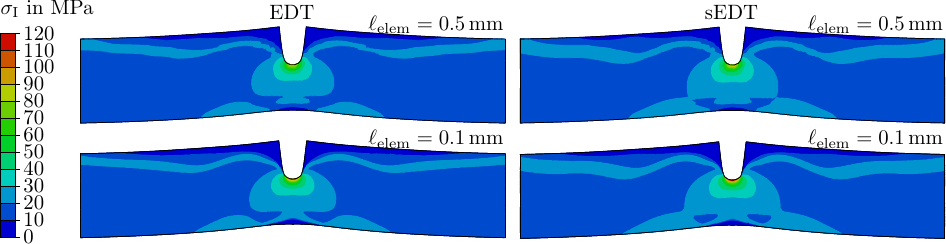}
	\caption{Maximum principal stress $\sigma_{\mathrm{I}}$ around the notch at the time $t=\qty{3}{\second}$ for the coarsest and finest mesh sizes considered with displacements unscaled.}
	\label{fig:sent_plot_contour_sMaxPr}
\end{figure}

Fig.~\ref{fig:sent_plot_contour_sMaxPr} shows the maximum principal stress around the notch. At its root, the peak stress reaches about \qtyrange{95}{100}{\mega\pascal} on the coarsest mesh, which does not resolve the notch radius, and about \qtyrange{115}{122}{\mega\pascal} on the finest mesh, with sEDT about \qty{5}{\percent} above EDT on both meshes. The stress fields of both models agree around the notch root. Towards the lower edge, the zone of moderate stress of sEDT extends further than that of EDT.

\begin{figure}[ht]
	\centering
	\captionsetup[subfigure]{font=normalsize}
	\includegraphics{plots/sent/sent-legend.pdf}\\[0.8em]
	\begin{subfigure}[t]{0.22\textwidth}
		\centering
		\includegraphics[width=\textwidth]{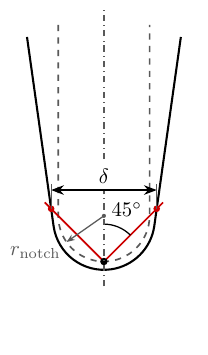}
		\caption{construction}
		\label{fig:sent_notch_opening}
	\end{subfigure}
	\hfill
	\begin{subfigure}[t]{0.37\textwidth}
		\centering
		\includegraphics[width=\textwidth]{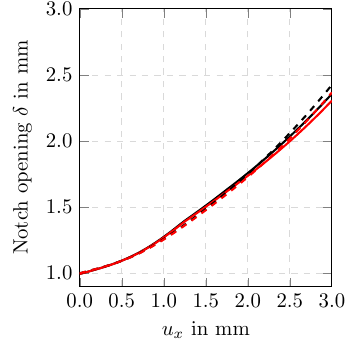}
		\caption{notch opening}
		\label{fig:sent_plot_ctod}
	\end{subfigure}
	\hfill
	\begin{subfigure}[t]{0.37\textwidth}
		\centering
		\includegraphics[width=\textwidth]{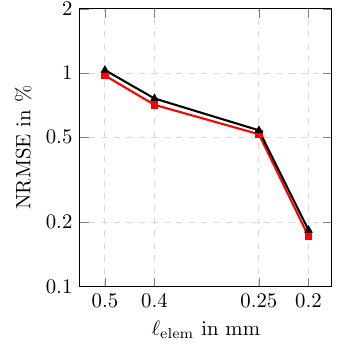}
		\caption{error under spatial refinement}
		\label{fig:sent_plot_nrmse_ctod}
	\end{subfigure}
	\caption[Notch opening of the SENT specimen]{Notch opening of the SENT specimen, measured along rays at \qty{\pm45}{\degree} from the initial notch root to the deformed flanks (solid, undeformed notch dashed), over the prescribed displacement on two meshes and its error~\eqref{eq:numerics_setup_error} under spatial refinement, each model against its own response on the finest mesh.}
	
	\label{fig:sent_plot_ctod_error}
\end{figure}

Fig.~\ref{fig:sent_plot_ctod} shows the notch opening $\delta$ over the prescribed displacement. Adapting the \qty{90}{\degree} intercept construction of Shih~\cite{Shih1981} to a blunt notch, we measure the opening $\delta$ as the distance between the two points at which lines drawn at \qty{\pm45}{\degree} to the notch axis from the initial position of the notch root intersect the deformed notch flanks, see Fig.~\ref{fig:sent_notch_opening}. The opening $\delta$ includes the bending of the strip, as the rays start from the initial position of the notch root. We therefore use it only to compare the two models on identical meshes, not as a fracture-mechanical quantity. Initially, the opening $\delta$ equals the notch width of \qty{1}{\milli\metre}. The openings of both models agree closely on all meshes, and their error decreases under mesh refinement from about \qty{1}{\percent} to \qty{0.2}{\percent}, see Fig.~\ref{fig:sent_plot_nrmse_ctod}. At the end of the load path, EDT opens the notch by about \qty{2.4}{\percent} more than sEDT, which conforms with its stronger softening.

\begin{figure}[ht]
	\centering
	\captionsetup[subfigure]{font=normalsize}
	\begin{subfigure}[t]{0.48\textwidth}
		\centering
		\includegraphics[width=\textwidth]{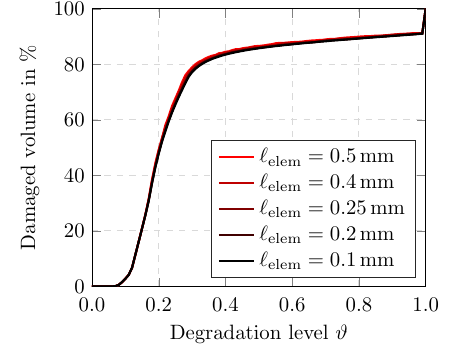}
		\caption{at the end of the load path}
		\label{fig:sent_plot_fBar_dmgvol}
	\end{subfigure}
	\hspace{1em}
	\begin{subfigure}[t]{0.48\textwidth}
		\centering
		\includegraphics[width=\textwidth]{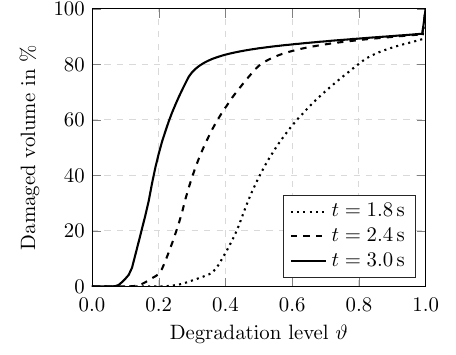}
		\caption{at three instants}
		\label{fig:sent_plot_fBar_dmgVol_times}
	\end{subfigure}
	\caption[Degraded volume fraction of the SENT specimen for sEDT]{Fraction of the specimen volume whose degradation $f$ is at most the level $\vartheta$, for sEDT, at the end of the load path for every element size and on the finest mesh, with $\ell_{\mathrm{elem}}=\qty{0.1}{\milli\metre}$, at the three instants of Fig.~\ref{fig:sent_plot_contour_fBar}.}
	
	\label{fig:sent_plot_fBar}
\end{figure}

Fig.~\ref{fig:sent_plot_fBar} quantifies how far the degradation of sEDT spreads. For a level $\vartheta$, it shows the fraction of the specimen volume in which the degradation $f$ is at most $\vartheta$, i.e., in which the stiffness dropped to at most the fraction $\vartheta$ of its undamaged value. We evaluate this fraction over the integration points, each weighted with the volume it represents. At the end of the load path, this volume fraction is almost identical on all meshes over the whole range of the level $\vartheta$, see Fig.~\ref{fig:sent_plot_fBar_dmgvol}. No localization is visible, as a localized band would narrow with the element size, so that the degraded volume fraction would decrease under mesh refinement. On the finest mesh, the curves shift towards lower levels as the load increases, see Fig.~\ref{fig:sent_plot_fBar_dmgVol_times}, i.e., the degradation decreases throughout the specimen and not only at the notch. At the end of the load path, the degradation dropped to at most $0.2$ in half of the specimen volume.

\begin{figure}[ht]
	\centering
	\includegraphics[width=\textwidth]{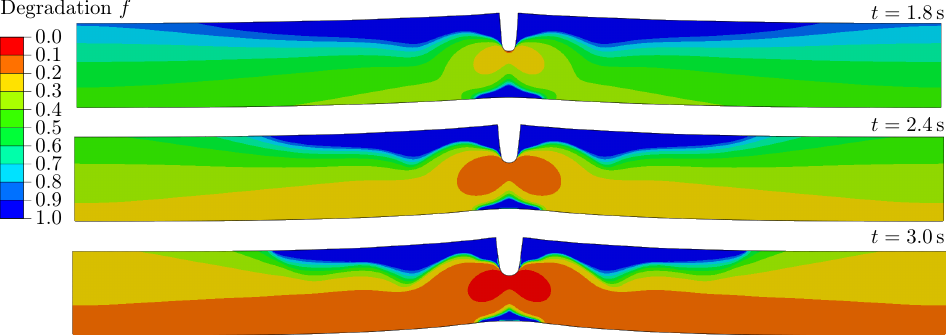}
	\caption{The degradation $f$ of sEDT on the mesh with $\ell_{\mathrm{elem}}=\qty{0.1}{\milli\metre}$ at three instants of the load path.}
	\label{fig:sent_plot_contour_fBar}
\end{figure}

Fig.~\ref{fig:sent_plot_contour_fBar} shows the degradation fields at the same three instants. The material degrades most in two lobes below the notch root, which grow as the load progresses. Away from the notch, the degradation decreases from the upper towards the lower edge, as the eccentric ligament bends the specimen. The material next to the traction-free notch flanks remains intact.

\subsection{A jet-engine bracket at industrial resolution}
\label{sec:numerics_bracket}

Finally, we consider a challenging industrial component of complex geometry, the bracket number $270$ of the SimJEB data set~\cite{Whalen2021}, which is made of Ti-6Al-4V. It is a lifting bracket that is bolted to the casing of a jet engine through its four holes and receives a pin through its fork. It carries the weight of the engine while the engine is handled on the ground. The elastic constants are listed in Tab.~\ref{tab:numerics_setup_materials}, but the data set provides no damage parameters. We therefore identify them from the stress limit $\sigma_c$ of \qty{903.2}{\mega\pascal}  prescribed by the original design challenge~\cite{Carter2014}. According to the evolution law~\eqref{eq:theory_scalar_model_evolution}, damage initiates once the driving force~\eqref{eq:theory_scalar_model_driving_force} exceeds the sum $r + \lambda$, and for the undamaged body and the exponential degradation this force equals the elastic energy density. Evaluating it for a state of uniaxial tension at the stress limit hence  fixes the sum,
\begin{equation}
	\label{eq:numerics_bracket_threshold}
	r + \lambda = \frac{\sigma_c^2}{2E} = \qty{3.584}{\mega\pascal}
	\qquad \text{and} \qquad
	\lambda = 0.1 \, r ,
\end{equation}
where the second relation splits the sum into the two parameters, whose values are listed in Tab.~\ref{tab:numerics_setup_materials}. Under uniaxial tension, damage thus initiates exactly at the limit value $\sigma_c$ of the original design challenge. The split and the damage-rate limitation, in contrast, are assumed, as  the data set provides no information on damage evolution. The computation therefore demonstrates the robustness and the scalability of the scalar update at industrial resolution.

\begin{figure}[ht]
	\centering
	\includegraphics[width=0.979\textwidth]{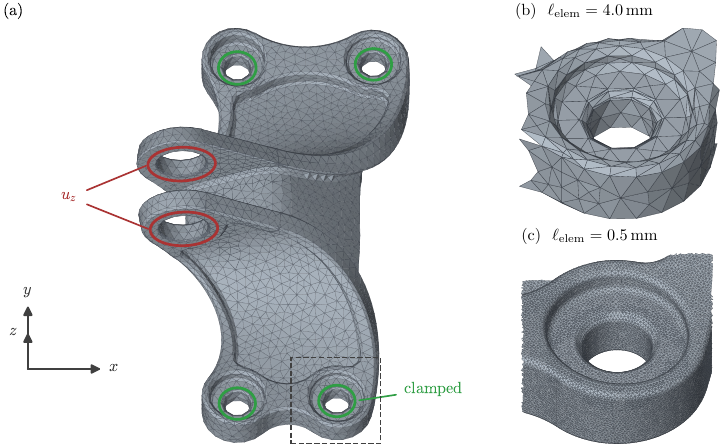}
	\caption{Bracket number 270 of the SimJEB data set~\cite{Whalen2021}. (a) The body, the bolt holes marked in green being clamped and the wall of the interface bore marked in red. (b), (c) The dashed detail at the coarsest and at the finest discretization of the mesh study.}
	\label{fig:bracket_geometry}
\end{figure}

The walls of the four bolt holes shown in Fig.~\ref{fig:bracket_geometry} are clamped, and the wall of the interface bore is displaced by \qty{8}{\milli\metre} in the $z$-direction within the load duration $T$ of \qty{1.125}{\second}, divided into $300$ equal load steps. The bore passes through both lugs of the fork, so its wall consists of two bands, each \qty{7.3}{\milli\metre} wide. Both bands are displaced, as the pin would load both lugs in the real assembly. Together they form a single Dirichlet boundary condition, and the reaction force $F$ reported below is accordingly the sum of the nodal reaction forces over the entire bore wall in the direction of the prescribed displacement.

Without further measures, this load introduction would damage the titanium directly at the hole walls, whereas in the real assembly the load is transferred there through steel fasteners and a steel pin. We model their effect by a collar within \qty{2}{\milli\metre} of every hole wall, in which the damage-initiation threshold $r$ and the entropic regularization parameter $\lambda$ are raised by a factor of three. As the initiation stress scales with the square root of the threshold, the material of the collar initiates damage under uniaxial tension at \qty{1564}{\mega\pascal}, which is $\sqrt{3}$ times the limit $\sigma_c$. The collar thus ensures that failure occurs within the structure rather than at the boundary conditions.

The computations of this subsection rely on a finite-element implementation of sEDT in DOLFINx~\cite{Baratta2023}. The meshes are generated by Gmsh~\cite{Geuzaine2009} with a uniform target element size and without curvature-based refinement, and neither the mesh nor the time step size is adapted during the computation. The displacement field is discretized by quadratic tetrahedra, and the constitutive state is stored at four quadrature points per element.

As the bracket undergoes large rotations under the prescribed displacement, we use a total Lagrangian formulation, in which the constitutive update receives the Green--Lagrange strain in place of the linearized strain and returns the second Piola--Kirchhoff stress. The joint convexity of the incremental potential in the strain and the reciprocal degradation  holds with the Green--Lagrange strain as argument, but not with respect to the displacement, on which this strain depends nonlinearly. The same applies to the remaining material-point results of Sec.~\ref{sec:theory}, whereas the positive definiteness of the global tangent is no longer guaranteed. The damage update of Sec.~\ref{sec:theory_frozen_metric} is local to each quadrature point and enters the global problem only through the degraded stiffness.

The global equilibrium equations are solved by Newton's method~\cite{NocedalWright2006}, terminated once the residual norm drops below $10^{-8}$ of its value at the beginning of the load step. The linear systems are solved by the conjugate-gradient method, preconditioned by smoothed-aggregation algebraic multigrid~\cite{Vanek1996}, with the six rigid-body modes as the near-nullspace. Each computation runs on $64$ MPI ranks, one per physical core of the dual-socket workstation described in Sec.~\ref{sec:numerics_setup}.

\begin{figure}[ht]
	\centering
	\includegraphics[width=442.77pt]{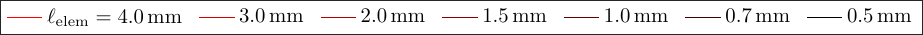}
	\\[0.8em]
	\begin{subfigure}[t]{0.48\textwidth}
		\centering
		\includegraphics[width=\textwidth]{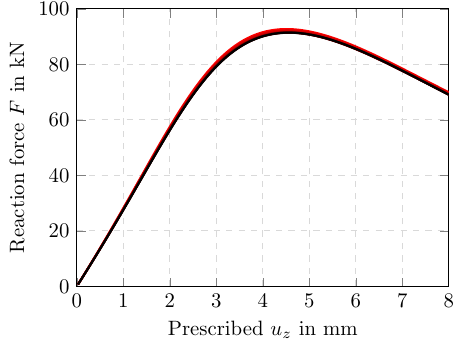}
		\caption{load path}
		\label{fig:bracket_response_loadpath}
	\end{subfigure}
	\hfill
	\begin{subfigure}[t]{0.48\textwidth}
		\centering
		\includegraphics[width=\textwidth]{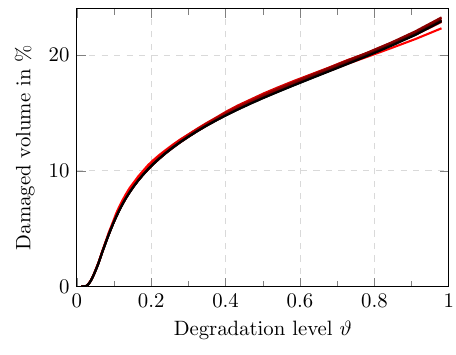}
		\caption{damaged volume}
		\label{fig:bracket_response_volume}
	\end{subfigure}
	\caption{Mesh study of the bracket, one curve per element size. (a) Reaction force at the interface bore over the prescribed displacement. (b) Fraction of the volume of the body outside the collars whose degradation is at most the level $\vartheta$ at the end of the load path.}
	\label{fig:bracket_response}
\end{figure}

We consider seven meshes of the same geometry, with element sizes ranging from \qty{4.0}{\milli\metre} down to \qty{0.5}{\milli\metre}. The finest mesh comprises more than three million elements and \num{13.1} million quadrature points, at each of which the constitutive update is performed. The load path in Fig.~\ref{fig:bracket_response_loadpath} rises to a peak load $F_{\max}$ of \qty{91.45}{\kilo\newton}, attained at the prescribed displacement $u^\ast$ of \qty{4.56}{\milli\metre}, and subsequently softens by \qty{24.48}{\percent} by the end of the path. The curves of all seven meshes are indistinguishable at the scale of the figure. Even the coarsest mesh reproduces the initial stiffness $K_0$ of the finest mesh to within \qty{3.12}{\percent}, and both the peak and the final force to within \qty{1.40}{\percent}. From the element size \qty{1.5}{\milli\metre} onwards, both reaction forces differ from those of the finest mesh by less than \qty{0.1}{\percent}. Interestingly, the error in the initial stiffness $K_0$, which is evaluated on the undamaged body and thus reflects the elastic discretization alone, is the largest of the three on every mesh. This suggests that the damage evolution adds little discretization error beyond that of the elastic problem on the meshes at hand, although all errors are measured against the finest mesh, which is not converged itself. Tab.~\ref{tab:bracket_ladder} summarizes these results together with the runtime and memory requirements of each computation. From the coarsest to the finest mesh, the number of elements grows by a factor of $115$, the runtime by a factor of $198$ and the memory by a factor of eight. On all seven meshes, the global Newton method converged in every load step without a reduction of the step size, and neither a local update nor a linear solve failed.

\begin{table}[ht]
	\centering
	\caption{The mesh study of the bracket. $K_0$ is the initial stiffness of the undamaged body, $F_{\max}$ the peak reaction force and $u^\ast$ the prescribed displacement at which it is attained, $F_8$ the reaction force at the end of the load path and $f_{\min}$ the smallest degradation attained in the body. Runtime and memory refer to the whole computation on all MPI ranks, the memory being the sum over all ranks.}
	\label{tab:bracket_ladder}
	\small
	\setlength{\tabcolsep}{0pt}
	\begin{tabular*}{\textwidth}{@{\extracolsep{\fill}}cccccccccc}
		\toprule
		$\ell_{\mathrm{elem}}$ & $n_{\mathrm{el}}$ & $n_{\mathrm{dof}}$ & runtime & memory & $K_0$ & $F_{\max}$ & $u^\ast$ & $F_8$ & $f_{\min}$\\
		{[\unit{\milli\metre}]} & [--] & [--] & [\unit{\second}] & [\unit{\giga\byte}] & [\unit{\kilo\newton\per\milli\metre}] & [\unit{\kilo\newton}] & [\unit{\milli\metre}] & [\unit{\kilo\newton}] & [--]\\
		\midrule
		$4.0$ & \num{28255} & \num{153678} & \num{117} & $10.63$ & $27.33$ & $92.73$ & $4.51$ & $69.96$ & $0.0142$\\
		$3.0$ & \num{43201} & \num{233106} & \num{159} & $11.00$ & $27.12$ & $92.29$ & $4.53$ & $69.71$ & $0.0147$\\
		$2.0$ & \num{90647} & \num{473133} & \num{386} & $12.30$ & $26.88$ & $91.72$ & $4.53$ & $69.34$ & $0.0135$\\
		$1.5$ & \num{168098} & \num{843618} & \num{785} & $14.42$ & $26.79$ & $91.45$ & $4.53$ & $69.09$ & $0.0135$\\
		$1.0$ & \num{470473} & \num{2232537} & \num{2720} & $21.94$ & $26.58$ & $91.50$ & $4.56$ & $69.11$ & $0.0127$\\
		$0.7$ & \num{1252022} & \num{5686632} & \num{7797} & $39.89$ & $26.54$ & $91.39$ & $4.56$ & $69.05$ & $0.0128$\\
		$0.5$ & \num{3263055} & \num{14333466} & \num{23153} & $84.80$ & $26.50$ & $91.45$ & $4.56$ & $69.06$ & $0.0127$\\
		\bottomrule
	\end{tabular*}
\end{table}

Over the investigated range of element sizes, the softening of this component shows almost no mesh dependence, see Fig.~\ref{fig:bracket_response_volume}. If damage localized into a single layer of elements, the damaged volume would scale with the element size. As the coarsest element size is eight times the finest, the damaged volume would then decrease by a factor of eight across the study. Instead, the seven curves lie on top of one another, apart from a slight deviation of the coarsest mesh above $\vartheta=0.8$, which stems from the discretization error of the structure. The volume fraction, evaluated outside the collars at the hole walls, with a degradation below \num{0.607}, which corresponds to a damage of one half, is \qty{17.78}{\percent} on the coarsest mesh and \qty{17.68}{\percent} on the finest, and down to a degradation of \num{0.05}, the fractions of any two meshes differ by at most \qty{5.2}{\percent}. Only below this level, which about \qty{1.4}{\percent} of the body reaches, does the relative spread increase. As the model carries no internal length, this observation is confined to the component and load path at hand.

\begin{figure}[ht]
	\centering
	\begin{subfigure}[b]{0.28\textwidth}
		\centering
		\includegraphics[width=\textwidth]{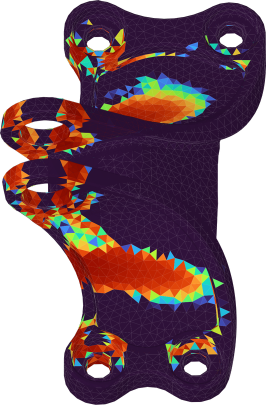}
		\caption{degradation $f$, element size $\ell_{\mathrm{elem}} = \qty{4.0}{\milli\metre}$}
		\label{fig:bracket_fields_coarse}
	\end{subfigure}
	\hfill
	\begin{subfigure}[b]{0.28\textwidth}
		\centering
		\includegraphics[width=\textwidth]{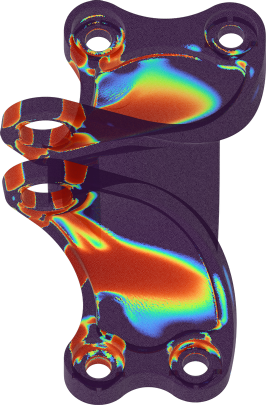}
		\caption{degradation $f$, element size $\ell_{\mathrm{elem}} = \qty{0.5}{\milli\metre}$}
		\label{fig:bracket_fields_fine}
	\end{subfigure}
	\hfill
	\begin{subfigure}[b]{0.28\textwidth}
		\centering
		\includegraphics[width=\textwidth]{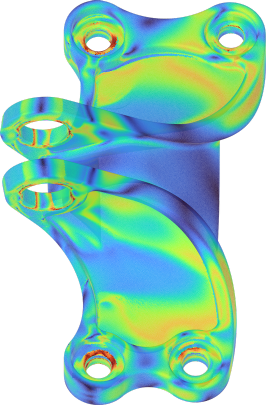}
		\caption{von Mises stress $\sigma_{\mathrm{vM}}$, element size $\ell_{\mathrm{elem}} = \qty{0.5}{\milli\metre}$}
		\label{fig:bracket_fields_stress}
	\end{subfigure}
	\\[0.2em]
	\includegraphics[width=\textwidth]{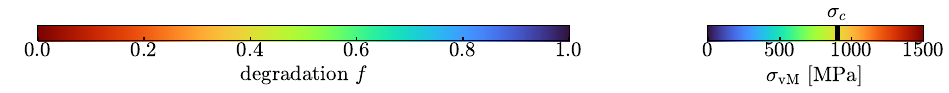}
	\caption{Degradation and von Mises stress at the end of the load path for two element sizes, each field drawn on the mesh of the computation it comes from. The black mark on the stress bar is the limit $\sigma_c$ from which the parameters $r$ and $\lambda$ of this component are derived by Eq.~\eqref{eq:numerics_bracket_threshold}.}
	\label{fig:bracket_fields}
\end{figure}

Fig.~\ref{fig:bracket_fields} shows the degradation on the coarsest and finest meshes, which locate the damage almost identically. Damage concentrates along the curved web that transfers the load from the interface fork to the lower pair of bolt holes, and around the five holes just outside their collars. The coarse mesh in Fig.~\ref{fig:bracket_fields_coarse} resolves this band element by element, so its outline is correspondingly irregular. Nevertheless, the band is connected rather than scattered over isolated elements, it lies where the fine mesh in Fig.~\ref{fig:bracket_fields_fine} places it, and it reaches the same damage level. This agreement is the field counterpart of the volume fractions reported above.

Fig.~\ref{fig:bracket_fields_stress} shows the von Mises stress of the same computation and reveals the effect of the degradation on the band. On the part of the surface carrying a damage above \num{0.5}, the median von Mises stress is \qty{435}{\mega\pascal}, compared with \qty{522}{\mega\pascal} on the remainder of the surface, and only \qty{3}{\percent} of the damaged surface still exceeds the stress limit $\sigma_c$. Where the damage exceeds two, the median decreases further to \qty{365}{\mega\pascal}. The band transfers its load to the surrounding material rather than accumulating stress, which is the local manifestation of the global softening in Fig.~\ref{fig:bracket_response_loadpath}. For a higher accuracy, the model would have to account for the ductile failure of Ti-6Al-4V by void growth and coalescence~\cite{Lecarme2014,RuizDeSotto2022}.

\newpage

\section{Conclusions}
\label{sec:conclusion}

Assessing the load-carrying capacity and the lifetime of components as well as structures is a central task of engineering practice. It is dissatisfying that the softening damage models commonly used for this purpose do not lead to well-posed boundary value problems, a deficiency which is typically absorbed by safety factors. Engineering practice accepts this state of affairs, but it does not have to.\\
The work at hand is part of a line of research which aims at mathematically well-posed damage models. Its central idea goes back to Schwarz et al.~\cite{SJH}, who combined a limitation of the damage rate with a convex relaxation technique. Lauff et al.~\cite{Lauff2026_OTModel} carried out this relaxation analytically and wrote down the associated model, a critical step towards practical use, as it keeps the computational effort at the material point manageable. The resulting model, however, does not lead to a strictly convex condensed incremental potential~\cite[Sec.~C.3]{Lauff2026_OTModel}. In particular, uniqueness of solutions to the associated mechanical boundary value problems is not directly inferred. As a remedy, the subsequent work~\cite{Lauff2026_OTRModel} introduced an entropic regularization which preserves the underlying mechanical structure, i.e., the model remains a generalized standard material and, with the threshold adjusted accordingly, its elastic domain remains unchanged. In this way, a well-posed model was obtained. Its solutions are measure-valued, as expected for relaxed nonconvex problems~\cite{Carstensen2002}. In practice, this circumstance is unfavorable, as the number of degrees of freedom per material point grows without bound as the time step size tends to zero. Solving a single scalar equation per material point~\cite{Schneider2026FastOTR} and redistributing onto finitely many damage states~\cite{Schneider2026_Redistribution} mitigated this drawback.\\
This work started off with a radical step. We studied the continuous-time limit of Entropic Damage Transport to understand which profiles of damage distributions are optimal. Indeed, the EDT model rested on a fixed discretization in time, an implicit Euler step, followed by a relaxation, so that everything was tied to this specific discretization in time. Surprisingly, in the time-continuous limit, a damage distribution that is concentrated at a single state initially remains concentrated throughout its evolution. In particular, it may be described by a single number, not a full measure. This result provides a retrospective explanation why the dedicated solver~\cite{Schneider2026FastOTR} and the redistribution~\cite{Schneider2026_Redistribution} performed so well.  Moreover, it reveals the damage distribution of Entropic Damage Transport as an artifact of the time discretization, namely of an upwind scheme for a transport equation in the space of damage states, see Eq.~\eqref{eq:apx_continuous_limit_mass_update}.\\
Thus, the question arose whether the scalar evolution admits a structure-preserving time discretization. The first step was a change of coordinates to the reciprocal degradation, in which the free energy becomes jointly convex~\cite{Goerthofer2022}. This step is natural and should find its way into damage mechanics, as it provides a canonical form of the free energy, similar to plasticity. In continuous time, the choice of the damage variable is thus a matter of convenience. Indeed, the structure of a generalized standard material is invariant under arbitrary changes of coordinates of its internal variables at the continuous level. This invariance does not carry over to the time discretization, however, and the price to pay for the convex free energy is the state dependence it induces in the dissipation.\\
Against this background, the frozen-metric discretization is a natural step. It retains the properties of the measure-valued model, such as a strongly convex condensed incremental potential and a well-posed incremental boundary value problem. In retrospect, the relaxation was not needed to obtain a well-posed model. Schwarz et al.~\cite{SJH} could have retained the scalar damage variable and chosen a more suitable time discretization instead. In a sense, the circle closes here. We returned to a scalar damage model in the spirit of the original damage mechanics of Kachanov~\cite{Kachanov1999}, and convex damage mechanics is embedded consistently into the framework of generalized standard materials. With this step, the mathematical exploration at material-point level is completed, and the focus may shift to material modeling, for instance to the coupling with plasticity or chemical processes, to the asymmetry between tension and compression and to anisotropic damage.\\
The computational experiments confirm the practical benefit of the scalar update. On the fiber-reinforced unit cell, it reduced the runtime by up to a factor of $6.4$ and the memory of the internal variables by a factor of $117$ compared with the original Entropic Damage Transport. For a jet-engine bracket with more than $14$ million degrees of freedom, the global Newton method converged in every load step without a reduction of the step size. The frozen metric lets the damage lag behind the scalar evolution, and time discretizations of higher accuracy which retain the convexity of the increment are the subject of ongoing work.\\
Further mathematical questions remain at the structural level. Every damage model whose reciprocal degradation is convex and whose dissipation potential is convex in the damage rate may be transformed to the reciprocal degradation and discretized with a frozen metric, which leads to a convex, though not necessarily strictly convex, incremental problem. Convexity at material-point level does not imply that this scheme yields a meaningful time discretization for boundary value problems, however, and the conditions which ensure the latter remain to be identified.\\
A different aspect concerns the physical background. Nonlocal damage mechanics is commonly motivated as follows~\cite{Bazant1976,Bazant2002}. A rate-independent softening model concentrates the damage on a singular surface, and an internal length is introduced to prevent this concentration and to explain size effects at the same time. In a sense, this reasoning is inconsistent. Experiments typically consider a window of strain rates of interest, within which the relevant quantities, such as the strain at failure and the maximum stress, depend only weakly on the rate. This observation forms the basis for modeling the process as rate-independent. Rate independence is thus an idealization valid within this window, and, in heterogeneous materials in particular, it is not known in advance whether the local strain rates remain inside it. Upon softening, the local strain rates increase and leave this window, and the concentration on a singular surface, which proceeds arbitrarily fast in a rate-independent model, is a consequence of the idealization.\\
Physically, however, damage cannot evolve arbitrarily fast, as microcracks grow at a finite speed, bounded by the Rayleigh wave speed~\cite{Freund1990}. Rate-limited models account for this physical limit directly, at the price of a rate dependence. To what extent this rate dependence matches experiments is open to date. This question is linked to the length scale, which arises naturally in real materials, on the one hand through heterogeneities such as the aggregates in concrete or the reinforcing particles of composites, and on the other hand through heterogeneities of the strength, as observed in experiments on the size effect~\cite{Weibull1951}. Translated to the model at hand, the parameters $r$ and $\lambda$ might themselves be heterogeneous with a certain correlation length, which could explain the size effect.

\section*{CRediT authorship contribution statement}

\textbf{Jonas Nutz:} Conceptualization, Methodology, Formal analysis, Software, Investigation, Writing -- original draft, Writing -- review \& editing. \textbf{Vikram Laheri:} Software, Investigation, Writing -- review \& editing. \textbf{Matti Schneider:} Conceptualization, Methodology, Formal analysis, Supervision, Writing -- review \& editing, Funding acquisition.

\section*{Declaration of competing interest}

The authors declare that they have no known competing financial interests or personal relationships that could have appeared to influence the work reported in this paper.

\section*{Funding}

This work was funded by the European Union through the European Research Council under the Horizon Europe programme (ERC Starting Grant BeyondRVE, grant agreement No.~101040238). Views and opinions expressed are, however, those of the authors only and do not necessarily reflect those of the European Union or the European Research Council Executive Agency. Neither the European Union nor the granting authority can be held responsible for them.

\section*{Data availability}

The data that support the findings of this study are available from the corresponding author upon reasonable request. An Abaqus user material (\texttt{UMAT}) of the scalar Entropic Damage Transport (sEDT) model is available at \url{https://git.uni-due.de/publicsoftwareingmath/sedt-umat}, together with an example, a driver at material-point level and tests.

\section*{Declaration of generative AI and AI-assisted technologies in the writing process}

During the preparation of this work, the authors used Claude (Anthropic) in order to improve the language. After using this tool, the authors reviewed and edited the content as needed and take full responsibility for the content of the publication.

\appendix 

\section{Convergence of the time-discrete Damage-Transport models}
\label{apx:convergence}

\subsection{Continuous-time limit of Entropic Damage Transport}
\label{apx:continuous_limit}

We show that the effective degradation of Entropic Damage Transport~\cite{Lauff2026_OTRModel}, started from a single damage state $d_0$, converges to the degradation along the scalar evolution~\eqref{eq:theory_scalar_model_evolution}. On the uniform time grid
\begin{equation} \label{eq:theory_scalar_model_time_grid}
	\triangle t:=\frac{T}{n_{\triangle t}} \quad\text{and}\quad t^n:=n \, \triangle t, \qquad n=0,1,\ldots,n_{\triangle t}.
\end{equation}
let $\bar f_{\triangle t}$ denote its piecewise constant interpolation and $d$ the solution of Eq.~\eqref{eq:theory_scalar_model_evolution} with $d(0)=d_0$. Suppose that $\lambda>0$, that the reciprocal degradation satisfies the assumptions~\eqref{eq:theory_scalar_model_reciprocal_degradation_boundary} and~\eqref{eq:theory_scalar_model_regularities}, and that the strain path and the initial state satisfy
\begin{equation} \label{eq:theory_scalar_model_strain_path_regularity}
	\feps\in W^{1,\infty}(0,T;\Sym{3}),
\end{equation}
and
\begin{equation} \label{eq:theory_scalar_model_attainable_damage_range}
	d_0 + T\ddotmax < \dmax.
\end{equation}
Then, there is a constant $c>0$, independent of the number of load steps $n_{\triangle t}$, such that the estimate
\begin{equation} \label{eq:theory_scalar_model_continuous_limit_estimate}
	\max_{0\leq n\leq n_{\triangle t}} \left| \bar f_{\triangle t}(t^n)-f(d(t^n)) \right| \leq c\sqrt{\triangle t}
\end{equation}
and thus the uniform convergence
\begin{equation} \label{eq:theory_scalar_model_continuous_limit}
	\bar f_{\triangle t}\longrightarrow f(d) \quad\text{uniformly on }[0,T] \quad\text{as}\quad n_{\triangle t} \rightarrow \infty
\end{equation}
hold.

\begin{figure}[ht]
	\centering
	\captionsetup[subfigure]{font=normalsize}
	\begin{subfigure}[b]{0.32\textwidth}
		\centering
		\includegraphics{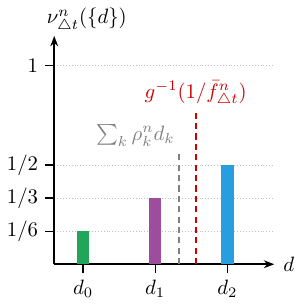}
		\caption{damage distribution}
		\label{fig:apx_continuous_limit_distribution_damage}
	\end{subfigure}
	\hfill
	\begin{subfigure}[b]{0.32\textwidth}
		\centering
		\includegraphics{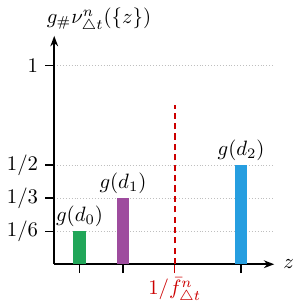}
		\caption{transformed distribution}
		\label{fig:apx_continuous_limit_distribution_image}
	\end{subfigure}
	\hfill
	\begin{subfigure}[b]{0.32\textwidth}
		\centering
		\includegraphics{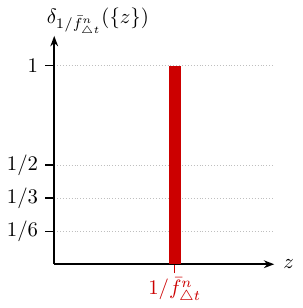}
		\caption{equivalent scalar state}
		\label{fig:apx_continuous_limit_distribution_scalar}
	\end{subfigure}
	\caption{Damage distribution and scalar state of equal free energy at one instant, for the exponential degradation~\eqref{eq:theory_scalar_model_exponential_degradation}.}
	
	\label{fig:apx_continuous_limit_distribution}
\end{figure}

We bound the quadratic deviation of the damage distribution from the exact solution by a recursion over the time steps and translate this bound into the estimate for the effective degradation. We recall the time-discrete Entropic Damage-Transport evolution in the notation required for the arguments, with the time index as a superscript, see Lauff et al.~\cite{Lauff2026_OTRModel} and Schneider~\cite{Schneider2026_Redistribution}. On the time grid~\eqref{eq:theory_scalar_model_time_grid}, we define the damage increment
\begin{equation}\label{eq:apx_continuous_limit_damage_increment}
	\triangle d := \triangle t \, \ddotmax .
\end{equation}
Starting from the initial damage state $d_0$, the attainable damage states are
\begin{equation}\label{eq:apx_continuous_limit_damage_states}
	\hat d_k := d_0 + k \triangle d, \qquad k = 0,1,\ldots,n_{\triangle t}.
\end{equation}
The damage distribution at time $t^n$ is a probability measure on the attainable states up to the index $n$,
\begin{equation}\label{eq:apx_continuous_limit_damage_distribution}
	\nu_{\triangle t}^n := \sum_{k=0}^n \rho_k^n \delta_{\hat d_k}, \qquad \rho_k^n \geq 0, \qquad \sum_{k=0}^n \rho_k^n = 1,
\end{equation}
with the weights $\rho_k^n$. The initial damage distribution is concentrated at the initial damage state, from which the exact solution starts as well,
\begin{equation}\label{eq:apx_continuous_limit_initial_condition}
	\rho_0^0 = 1 \quad\text{and}\quad d(0) = d_0 .
\end{equation}
The effective degradation is the reciprocal of the mean reciprocal degradation,
\begin{equation}\label{eq:apx_continuous_limit_effective_degradation}
	\bar f_{\triangle t}^n := \left[\sum_{k=0}^n \rho_k^n g(\hat d_k)\right]^{-1}.
\end{equation}
Fig.~\ref{fig:apx_continuous_limit_distribution} links the damage distribution~\eqref{eq:apx_continuous_limit_damage_distribution} of Entropic Damage Transport to the reciprocal degradation variable~\eqref{eq:theory_hidden_convexity_transformation}, the scalar internal variable of the frozen-metric formulation. The effective degradation~\eqref{eq:apx_continuous_limit_effective_degradation} depends on the distribution only through the mean of its image under the reciprocal degradation $g$, i.e., through the single value $1/\bar f_{\triangle t}^n$ of this variable. For every strain, the relaxed free energy $\bar f_{\triangle t}^n\,\freeEnergy_0(\feps)$ of the distribution, see Sec.~\ref{sec:theory_hidden_convexity}, thus coincides with the transformed free energy~\eqref{eq:theory_hidden_convexity_transformed_energy} of the scalar state at this value. The estimate~\eqref{eq:theory_scalar_model_continuous_limit_estimate} therefore compares this scalar state with the exact solution through its degradation, as the estimate~\eqref{eq:theory_frozen_metric_convergence_estimate} does for the frozen-metric states $z_n$.
The inverse function~\eqref{eq:theory_hidden_convexity_inverse} assigns to this value the damage $g^{-1}(1/\bar f_{\triangle t}^n)$, the counterpart of the frozen-metric damage states $d_n$ in Eq.~\eqref{eq:theory_frozen_metric_convergence_initial_state}. Jensen's inequality for the convex reciprocal degradation~\eqref{eq:theory_scalar_model_reciprocal_degradation}, combined with the monotonicity~\eqref{eq:theory_hidden_convexity_inverse_derivatives} of its inverse, bounds this damage from below by the mean damage of the distribution,
\begin{equation}\label{eq:apx_continuous_limit_jensen}
	g^{-1}\big(1/\bar f_{\triangle t}^n\big) = g^{-1}\left(\sum_{k=0}^n \rho_k^n g(\hat d_k)\right) \geq g^{-1}\left(g\left(\sum_{k=0}^n \rho_k^n \hat d_k\right)\right) = \sum_{k=0}^n \rho_k^n \hat d_k .
\end{equation}
As the function $g$ is strictly convex by the assumption following Eq.~\eqref{eq:theory_scalar_model_reciprocal_degradation}, equality holds only for a distribution concentrated at a single damage state. Fig.~\ref{fig:apx_continuous_limit_distribution_damage} marks both damages for a spread distribution. The time-discrete evolutions of Entropic Damage Transport and of the frozen-metric formulation differ in general for a positive time step size and approximate one and the same scalar evolution by the estimates~\eqref{eq:theory_scalar_model_continuous_limit_estimate} and~\eqref{eq:theory_frozen_metric_convergence_estimate}.

The transition from time $t^n$ to time $t^{n+1}$ is characterized implicitly through the damage volume fractions
\begin{equation}\label{eq:apx_continuous_limit_volume_fractions}
	\phi_k^{n+1} := \frac{\left\langle (\bar f_{\triangle t}^{n+1})^2 Q_{\triangle t}(\hat d_k) \freeEnergy_0(\feps(t^{n+1})) - r - \lambda \right\rangle_+}{(\bar f_{\triangle t}^{n+1})^2 Q_{\triangle t}(\hat d_k) \freeEnergy_0(\feps(t^{n+1})) - r}, \qquad k = 0,1,\ldots,n,
\end{equation}
where the fraction is defined to be zero whenever its numerator vanishes and where the secant of the reciprocal degradation reads
\begin{equation}\label{eq:apx_continuous_limit_secant_function}
	Q_{\triangle t}(d) := \frac{g(d + \triangle d) - g(d)}{\triangle d}, \qquad d,\, d+\triangle d \in \mathcal{D}.
\end{equation}
In one time step, the fraction $\phi_k^{n+1}$ of the weight at the state $\hat d_k$ moves to the neighboring state $\hat d_{k+1}$, whereas the remainder stays at the state $\hat d_k$,
\begin{equation}\label{eq:apx_continuous_limit_mass_update}
	\rho_0^{n+1} = (1 - \phi_0^{n+1}) \rho_0^n, \qquad \rho_k^{n+1} = (1 - \phi_k^{n+1}) \rho_k^n + \phi_{k-1}^{n+1} \rho_{k-1}^n, \qquad k = 1,2,\ldots,n,
\end{equation}
together with
\begin{equation}\label{eq:apx_continuous_limit_mass_update_last}
	\rho_{n+1}^{n+1} = \phi_n^{n+1} \rho_n^n .
\end{equation}
Collecting the terms of the updates~\eqref{eq:apx_continuous_limit_mass_update} and~\eqref{eq:apx_continuous_limit_mass_update_last} by their origin leads to the transport identity
\begin{equation}\label{eq:apx_continuous_limit_transport_identity}
	\sum_{j=0}^{n+1} \rho_j^{n+1} \omega(\hat d_j) = \sum_{k=0}^n \rho_k^n \left[(1 - \phi_k^{n+1}) \omega(\hat d_k) + \phi_k^{n+1} \omega(\hat d_{k+1})\right]
\end{equation}
for every function $\omega$ defined on the attainable damage states. The damage rate depends on the driving force through a single scalar function, which carries the entropic regularization. We define this kinetic function by
\begin{equation}\label{eq:apx_continuous_limit_kinetic_function}
	K(y) := \ddotmax\,\frac{\langle y-r-\lambda\rangle_+}{y-r} = \begin{cases}
		0, & y \leq r + \lambda,\\[1mm]
		\displaystyle \ddotmax \left(1 - \frac{\lambda}{y-r}\right), & y > r + \lambda,
	\end{cases}
\end{equation}
where the fraction is again defined to be zero whenever its numerator vanishes. The kinetic function is continuous, nondecreasing and bounded by the damage-rate limitation $\ddotmax$. It is constant on the inactive interval, whereas its derivative on the active interval obeys
\begin{equation}\label{eq:apx_continuous_limit_kinetic_derivative}
	0 < K'(y) = \frac{\ddotmax \lambda}{(y-r)^2} < \frac{\ddotmax}{\lambda}, \qquad y > r + \lambda,
\end{equation}
as the argument exceeds the threshold $r$ by more than the entropic regularization parameter $\lambda$ there. The bound is approached at the transition point $r+\lambda$. For two arguments on either side of the transition point, $y_1\leq r+\lambda<y_2$, the kinetic function vanishes at the first one and at the transition point, so that the mean-value theorem on the active interval gives
\begin{equation}\label{eq:apx_continuous_limit_kinetic_lipschitz_chain}
	\left|K(y_2)-K(y_1)\right| = \left|K(y_2)-K(r+\lambda)\right| \leq \frac{\ddotmax}{\lambda}\left(y_2-r-\lambda\right) \leq \frac{\ddotmax}{\lambda}\left(y_2-y_1\right).
\end{equation}
Hence, the mean-value theorem applied on either side of the transition point, together with continuity at this point and the chain~\eqref{eq:apx_continuous_limit_kinetic_lipschitz_chain}, leads to the global estimate
\begin{equation}\label{eq:apx_continuous_limit_kinetic_lipschitz}
	\left|K(y_1) - K(y_2)\right| \leq \frac{\ddotmax}{\lambda} |y_1-y_2|, \qquad y_1,y_2 \in \R.
\end{equation}
This is the only place where the entropic regularization parameter $\lambda$ enters the estimates. In terms of the kinetic function, the volume fractions~\eqref{eq:apx_continuous_limit_volume_fractions} are expressed as
\begin{equation}\label{eq:apx_continuous_limit_kinetic_volume_fractions}
	\ddotmax \phi_k^{n+1} = K\left((\bar f_{\triangle t}^{n+1})^2 Q_{\triangle t}(\hat d_k) \freeEnergy_0(\feps(t^{n+1}))\right), \qquad 0 \leq \phi_k^{n+1} < 1,
\end{equation}
and the evolution law~\eqref{eq:theory_scalar_model_evolution} with the driving force~\eqref{eq:theory_scalar_model_driving_force} takes the form
\begin{equation}\label{eq:apx_continuous_limit_scalar_evolution}
	\dot d(t) = F(t,d(t)), \qquad F(t,d) := K\left(\frac{g'(d)}{g(d)^2} \freeEnergy_0(\feps(t))\right).
\end{equation}
The transport identity~\eqref{eq:apx_continuous_limit_transport_identity} settles the implicit character of the volume fractions. Applied to the reciprocal degradation $g$, it turns the definition~\eqref{eq:apx_continuous_limit_effective_degradation} at time $t^{n+1}$ into the scalar equation
\begin{equation}\label{eq:apx_continuous_limit_transition_equation}
	\frac{1}{\bar f_{\triangle t}^{n+1}} = \sum_{k=0}^n \rho_k^n g(\hat d_k) + \triangle d \sum_{k=0}^n \rho_k^n Q_{\triangle t}(\hat d_k)\, \phi_k^{n+1}
\end{equation}
for the reciprocal of the new effective degradation. By Eq.~\eqref{eq:apx_continuous_limit_kinetic_volume_fractions}, this unknown enters the right-hand side only through the arguments of the kinetic function, which do not increase as the unknown grows. As the kinetic function is nondecreasing and the secant~\eqref{eq:apx_continuous_limit_secant_function} is positive, the right-hand side is continuous and nonincreasing in the unknown. The difference between the left-hand and the right-hand side thus increases strictly. It is nonpositive at the previous mean reciprocal degradation and positive at the value obtained by moving the entire weight to the neighboring states, as the volume fractions stay below one. Hence, Eq.~\eqref{eq:apx_continuous_limit_transition_equation} has exactly one solution. Inserting it into the volume fractions~\eqref{eq:apx_continuous_limit_kinetic_volume_fractions} and the updates~\eqref{eq:apx_continuous_limit_mass_update} and~\eqref{eq:apx_continuous_limit_mass_update_last} yields nonnegative weights of unit sum, and the transport identity~\eqref{eq:apx_continuous_limit_transport_identity} shows that this solution is the reciprocal of the resulting effective degradation. Hence, the transition exists and is unique. The attainable states~\eqref{eq:apx_continuous_limit_damage_states} occupied at time $t^n$ lie above the initial damage state by at most the product of the time $t^n$ and the damage-rate limitation, and the kinetic function bounds the rate of the exact solution by the damage-rate limitation. Thus, for every number of load steps, the exact solution and all attainable states lie in the interval
\begin{equation}\label{eq:apx_continuous_limit_interval}
	d(t),\, \hat d_k \in \mathcal{I} := \left[d_0,\, d_0 + T\ddotmax\right], \qquad t \in [0,T] \quad\text{and}\quad k = 0,1,\ldots,n_{\triangle t},
\end{equation}
which is a compact subset of the interval $\mathcal{D}$ by the condition~\eqref{eq:theory_scalar_model_attainable_damage_range}. The transition from time $t^n$ to time $t^{n+1}$ evaluates the secant~\eqref{eq:apx_continuous_limit_secant_function} between the neighboring states $\hat d_k$ and $\hat d_{k+1}$, both of which lie in the interval~$\mathcal{I}$. Throughout, the letter $c$ denotes a generic positive constant which may change from one estimate to the next. It depends only on the damage-rate limitation $\ddotmax$, the entropic regularization parameter $\lambda$, the maxima of the functions $g$, $g'$ and $g''$ on the interval $\mathcal{I}$, the maximum and the Lipschitz constant of the undamaged energy along the strain path~\eqref{eq:theory_scalar_model_strain_path_regularity}, the final time $T$ and the initial damage state $d_0$. In particular, every constant below is independent of the index $n$ and the time step size $\triangle t$. The function $F$ is Lipschitz continuous on $[0,T] \times \mathcal{I}$, as the kinetic function satisfies Eq.~\eqref{eq:apx_continuous_limit_kinetic_lipschitz}, the factor of the undamaged energy in the driving force~\eqref{eq:theory_scalar_model_driving_force} is continuously differentiable on the interval~$\mathcal{I}$, and the undamaged energy is bounded and Lipschitz continuous along the strain path. By the Picard--Lindel{\"o}f theorem, applied to the function $F$ with its second argument projected onto the interval~$\mathcal{I}$, and by the a priori bound~\eqref{eq:apx_continuous_limit_interval}, the exact solution exists, is unique and stays in the interval~$\mathcal{I}$. The fundamental theorem of calculus expresses the secant as the mean of the derivative $g'$ between two neighboring states. The boundedness of the function $g''$ on the interval~$\mathcal{I}$ thus implies the consistency estimate
\begin{equation}\label{eq:apx_continuous_limit_secant_consistency}
	\begin{aligned}
		\left|Q_{\triangle t}(\hat d_k) - g'(\hat d_k)\right| &= \frac{1}{\triangle d} \left| \int_{\hat d_k}^{\hat d_{k+1}} \left[g'(\xi) - g'(\hat d_k)\right] \diff \xi \right| \leq \frac{\|g''\|_{L^\infty(\mathcal{I})}}{\triangle d} \int_{\hat d_k}^{\hat d_{k+1}} (\xi - \hat d_k) \, \diff \xi\\
		&= \frac{1}{2} \|g''\|_{L^\infty(\mathcal{I})} \triangle d \leq c \triangle t, \qquad k = 0,1,\ldots,n_{\triangle t}-1.
	\end{aligned}
\end{equation}
The same representation bounds the secant by the maximum of the derivative $g'$ on the interval~$\mathcal{I}$.
We introduce the quadratic error
\begin{equation}\label{eq:apx_continuous_limit_quadratic_error}
	E^n := \sum_{k=0}^n \rho_k^n |\hat d_k-d(t^n)|^2.
\end{equation}
It measures the spread of the distribution around the exact solution, including the deviation of its mean. The free energy depends on the distribution only through the scalar state $1/\bar f_{\triangle t}^n$ of Fig.~\ref{fig:apx_continuous_limit_distribution}, whereas the volume fractions~\eqref{eq:apx_continuous_limit_volume_fractions} evaluate the secant at every damage state and thus depend on the whole distribution. The Cauchy--Schwarz inequality, applied to the square roots of the weights and the weighted deviations, and the normalization in Eq.~\eqref{eq:apx_continuous_limit_damage_distribution} imply the estimate
\begin{equation}\label{eq:apx_continuous_limit_first_error_moment}
	\sum_{k=0}^n \rho_k^n |\hat d_k-d(t^n)| \leq \left(\sum_{k=0}^n \rho_k^n\right)^{1/2} \left(\sum_{k=0}^n \rho_k^n |\hat d_k-d(t^n)|^2\right)^{1/2} = \sqrt{E^n}.
\end{equation}
We first control the effective degradation entering the transition from time $t^n$ to time $t^{n+1}$. Applying the transport identity~\eqref{eq:apx_continuous_limit_transport_identity} to the reciprocal degradation $g$ leads to
\begin{equation}\label{eq:apx_continuous_limit_reciprocal_transport}
	\sum_{j=0}^{n+1} \rho_j^{n+1} g(\hat d_j) = \sum_{k=0}^n \rho_k^n g(\hat d_k) + \sum_{k=0}^n \rho_k^n \phi_k^{n+1} \left[g(\hat d_{k+1}) - g(\hat d_k)\right].
\end{equation}
The reciprocal degradation is Lipschitz continuous on the interval~$\mathcal{I}$, and the volume fractions lie between zero and one by Eq.~\eqref{eq:apx_continuous_limit_kinetic_volume_fractions}. Thus, Eqs.~\eqref{eq:apx_continuous_limit_first_error_moment} and~\eqref{eq:apx_continuous_limit_reciprocal_transport} imply
\begin{equation}\label{eq:apx_continuous_limit_reciprocal_error}
	\begin{aligned}
		\left|\sum_{j=0}^{n+1} \rho_j^{n+1} g(\hat d_j) - g(d(t^n))\right| &\leq \sum_{k=0}^n \rho_k^n \left|g(\hat d_k) - g(d(t^n))\right| + \sum_{k=0}^n \rho_k^n \phi_k^{n+1} \left|g(\hat d_{k+1}) - g(\hat d_k)\right|\\
		&\leq c \sqrt{E^n} + c \triangle d \leq c \left(\sqrt{E^n} + \triangle t\right), \qquad n=0,1,\ldots,n_{\triangle t}-1.
	\end{aligned}
\end{equation}
Both the mean reciprocal degradation and the reciprocal degradation are at least one, so that their reciprocals differ by at most their difference. With the definition~\eqref{eq:apx_continuous_limit_effective_degradation}, the estimate~\eqref{eq:apx_continuous_limit_reciprocal_error} thus leads to
\begin{equation}\label{eq:apx_continuous_limit_new_degradation_error}
	\left|\bar f_{\triangle t}^{n+1} - f(d(t^n))\right| = \frac{\left|\sum_{j=0}^{n+1} \rho_j^{n+1} g(\hat d_j) - g(d(t^n))\right|}{\sum_{j=0}^{n+1} \rho_j^{n+1} g(\hat d_j) \; g(d(t^n))} \leq c \left(\sqrt{E^n} + \triangle t\right), \qquad n=0,1,\ldots,n_{\triangle t}-1.
\end{equation}
We define the discrete velocities
\begin{equation}\label{eq:apx_continuous_limit_discrete_velocities}
	v_k^{n+1} := K\left((\bar f_{\triangle t}^{n+1})^2 Q_{\triangle t}(\hat d_k) \freeEnergy_0(\feps(t^{n+1}))\right) = \ddotmax \phi_k^{n+1}, \qquad k=0,1,\ldots,n,
\end{equation}
which are the volume fractions scaled by the damage-rate limitation, see Eq.~\eqref{eq:apx_continuous_limit_kinetic_volume_fractions}. To compare them with the exact velocity, we decompose the arguments of the kinetic function. By the definition~\eqref{eq:theory_scalar_model_reciprocal_degradation}, the argument of the function~\eqref{eq:apx_continuous_limit_scalar_evolution} at the previous exact state carries the squared degradation times the derivative of the reciprocal degradation. The effective degradation and the degradation function take values in the interval $(0,1]$, so that the difference of their squares is at most twice their difference. Together with the boundedness of the secant, this leads to
\begin{equation}\label{eq:apx_continuous_limit_velocity_argument_error}
	\begin{aligned}
		\left|(\bar f_{\triangle t}^{n+1})^2 Q_{\triangle t}(\hat d_k)-f(d(t^n))^2 g'(d(t^n))\right| &\leq \left|(\bar f_{\triangle t}^{n+1})^2-f(d(t^n))^2\right| Q_{\triangle t}(\hat d_k) + f(d(t^n))^2 \left|Q_{\triangle t}(\hat d_k)-g'(d(t^n))\right|\\
		&\leq c\left|\bar f_{\triangle t}^{n+1}-f(d(t^n))\right|+\left|Q_{\triangle t}(\hat d_k)-g'(\hat d_k)\right|+\left|g'(\hat d_k)-g'(d(t^n))\right|.
	\end{aligned}
\end{equation}
The function $g'$ is Lipschitz continuous on the interval~$\mathcal{I}$, and the undamaged energy is bounded along the strain path. Thus, Eqs.~\eqref{eq:apx_continuous_limit_secant_consistency},~\eqref{eq:apx_continuous_limit_new_degradation_error} and~\eqref{eq:apx_continuous_limit_velocity_argument_error}, together with the Lipschitz estimate~\eqref{eq:apx_continuous_limit_kinetic_lipschitz}, lead to the pointwise bound
\begin{equation}\label{eq:apx_continuous_limit_velocity_error}
	\left|v_k^{n+1} - F(t^{n+1},d(t^n))\right| \leq c \left(|\hat d_k-d(t^n)| + \sqrt{E^n} + \triangle t\right), \qquad k=0,1,\ldots,n.
\end{equation}
Squaring the estimate~\eqref{eq:apx_continuous_limit_velocity_error}, bounding the square of the sum of three terms by three times the sum of their squares, multiplying by the weight $\rho_k^n$ and summing over the index $k$ result in
\begin{equation}\label{eq:apx_continuous_limit_squared_velocity_error}
	\sum_{k=0}^n \rho_k^n \left|v_k^{n+1} - F(t^{n+1},d(t^n))\right|^2 \leq c \left(E^n + (\triangle t)^2\right).
\end{equation}
We next derive a closed recursion for the quadratic error~\eqref{eq:apx_continuous_limit_quadratic_error}. With the exact increment
\begin{equation}\label{eq:apx_continuous_limit_exact_increment}
	\delta^n := d(t^{n+1})-d(t^n),
\end{equation}
applying the transport identity~\eqref{eq:apx_continuous_limit_transport_identity} to the squared distance from the exact state $d(t^{n+1})$ leads to
\begin{equation}\label{eq:apx_continuous_limit_error_transport}
	E^{n+1} = \sum_{k=0}^n \rho_k^n \left[(1 - \phi_k^{n+1}) |\hat d_k-d(t^n)-\delta^n|^2 + \phi_k^{n+1} |\hat d_k-d(t^n)+\triangle d-\delta^n|^2\right].
\end{equation}
The definitions~\eqref{eq:apx_continuous_limit_damage_increment} and~\eqref{eq:apx_continuous_limit_discrete_velocities} give the relation
\begin{equation}\label{eq:apx_continuous_limit_velocity_increment}
	\triangle d \, \phi_k^{n+1} = \triangle t \, v_k^{n+1}, \qquad k=0,1,\ldots,n.
\end{equation}
Expanding the second square in Eq.~\eqref{eq:apx_continuous_limit_error_transport} and inserting the relation~\eqref{eq:apx_continuous_limit_velocity_increment} lead to
\begin{equation}\label{eq:apx_continuous_limit_error_expansion}
	E^{n+1} = \sum_{k=0}^n \rho_k^n \left[|\hat d_k-d(t^n)-\delta^n|^2 + 2 \triangle t \, v_k^{n+1}(\hat d_k-d(t^n)-\delta^n) + (\triangle d)^2 \phi_k^{n+1}\right].
\end{equation}
Completing the square in the first two terms and using the relation~\eqref{eq:apx_continuous_limit_velocity_increment} once more, we find
\begin{equation}\label{eq:apx_continuous_limit_completed_square}
	\begin{aligned}
		E^{n+1} &= \sum_{k=0}^n \rho_k^n \left[|\hat d_k-d(t^n)-\delta^n+\triangle t \, v_k^{n+1}|^2 - (\triangle t \, v_k^{n+1})^2 + (\triangle d)^2 \phi_k^{n+1}\right]\\
		&= \sum_{k=0}^n \rho_k^n |\hat d_k-d(t^n)-\delta^n+\triangle t \, v_k^{n+1}|^2 + (\triangle d)^2 \sum_{k=0}^n \rho_k^n \phi_k^{n+1}\left(1-\phi_k^{n+1}\right).
	\end{aligned}
\end{equation}
This completion of the square is the central stability step. It retains the correlation between the transported state and its velocity, so that the spreading of the damage distribution and the error of its mean need not be estimated separately. The last term is the variance which the splitting of the weights produces in one time step. The product of a volume fraction and its complement is at most one quarter, so that the damage increment~\eqref{eq:apx_continuous_limit_damage_increment} leads to
\begin{equation}\label{eq:apx_continuous_limit_completed_square_bound}
	E^{n+1} \leq \sum_{k=0}^n \rho_k^n |\hat d_k-d(t^n)-\delta^n+\triangle t \, v_k^{n+1}|^2 + \frac{\ddotmax^2}{4}(\triangle t)^2.
\end{equation}
The exact increment satisfies the consistency relation
\begin{equation}\label{eq:apx_continuous_limit_exact_consistency}
	\delta^n = \triangle t \, F(t^{n+1},d(t^n)) + R^{n+1}, \quad R^{n+1} := \int_{t^n}^{t^{n+1}} \left[F(s,d(s))-F(t^{n+1},d(t^n))\right] \diff s, \quad |R^{n+1}| \leq c(\triangle t)^2.
\end{equation}
Integrating the evolution law~\eqref{eq:apx_continuous_limit_scalar_evolution} over the time step gives the representation of the exact increment. The Lipschitz continuity of the function $F$ bounds the integrand by the time step size and the damage change within the step, and the kinetic function bounds the latter by the product of the damage-rate limitation and the time step size,
\begin{equation}\label{eq:apx_continuous_limit_remainder_integrand}
	\left|F(s,d(s))-F(t^{n+1},d(t^n))\right| \leq c\left(|t^{n+1}-s|+|d(s)-d(t^n)|\right) \leq c\,\triangle t, \qquad s\in[t^n,t^{n+1}].
\end{equation}\\
We introduce the velocity defect
\begin{equation}\label{eq:apx_continuous_limit_velocity_defect}
	b_k^{n+1} := v_k^{n+1} - F(t^{n+1},d(t^n)), \qquad k=0,1,\ldots,n.
\end{equation}
Inserting the consistency relation~\eqref{eq:apx_continuous_limit_exact_consistency} and the velocity defect~\eqref{eq:apx_continuous_limit_velocity_defect} into the square in Eq.~\eqref{eq:apx_continuous_limit_completed_square_bound} leads to the identity

\begin{equation}\label{eq:apx_continuous_limit_square_identity}
	\hat d_k-d(t^n)-\delta^n+\triangle t \, v_k^{n+1} = \hat d_k-d(t^n)+\triangle t \, b_k^{n+1}-R^{n+1}, \qquad k=0,1,\ldots,n,
\end{equation}
which turns the bound~\eqref{eq:apx_continuous_limit_completed_square_bound} into
\begin{equation}\label{eq:apx_continuous_limit_error_before_young}
	E^{n+1} \leq \sum_{k=0}^n \rho_k^n |\hat d_k-d(t^n)+\triangle t \, b_k^{n+1}-R^{n+1}|^2 + c(\triangle t)^2.
\end{equation}
Young's inequality with the parameter $\triangle t/T$ reads

\begin{equation}\label{eq:apx_continuous_limit_young}
	|x+y|^2 \leq \left(1+\frac{\triangle t}{T}\right)|x|^2 + \left(1+\frac{T}{\triangle t}\right)|y|^2, \qquad x,y \in \R.
\end{equation}
The parameter is dimensionless, and as the time step size does not exceed the final time, the second Young factor is at most $2T/\triangle t$ for every number of load steps. The elementary estimate
\begin{equation}\label{eq:apx_continuous_limit_defect_remainder_split}
	|\triangle t \, b_k^{n+1}-R^{n+1}|^2 \leq 2(\triangle t)^2 |b_k^{n+1}|^2 + 2|R^{n+1}|^2
\end{equation}
together with the estimates~\eqref{eq:apx_continuous_limit_squared_velocity_error} and~\eqref{eq:apx_continuous_limit_exact_consistency} then leads to

\begin{equation}\label{eq:apx_continuous_limit_young_remainder_bound}
	\left(1+\frac{T}{\triangle t}\right)\sum_{k=0}^n \rho_k^n |\triangle t \, b_k^{n+1}-R^{n+1}|^2 \leq 4T\triangle t \sum_{k=0}^n \rho_k^n |b_k^{n+1}|^2 + \frac{4T|R^{n+1}|^2}{\triangle t} \leq c\triangle t \, E^n+c(\triangle t)^3.
\end{equation}
Applying Young's inequality~\eqref{eq:apx_continuous_limit_young} with the deviation $\hat d_k-d(t^n)$ as the first summand in Eq.~\eqref{eq:apx_continuous_limit_error_before_young} and inserting the bound~\eqref{eq:apx_continuous_limit_young_remainder_bound}, we arrive at the recursion
\begin{equation}\label{eq:apx_continuous_limit_error_recursion}
	E^{n+1} \leq \left(1+\frac{\triangle t}{T}\right) E^n + c\triangle t \, E^n + c(\triangle t)^3 + c(\triangle t)^2 \leq (1+c_2\triangle t) E^n + c_1(\triangle t)^2
\end{equation}
with the constants $c_2 := 1/T + c$ and $c_1 := c\,(1+T)$, where the inequality $(\triangle t)^3 \leq T(\triangle t)^2$ is used. The initial conditions~\eqref{eq:apx_continuous_limit_initial_condition} give a vanishing initial quadratic error $E^0$. Iterating the recursion~\eqref{eq:apx_continuous_limit_error_recursion}, summing the geometric series, bounding the growth factor by its exponential and using that the time $t^n$ does not exceed the final time $T$ lead to
\begin{equation}\label{eq:apx_continuous_limit_error_bound}
	\begin{aligned}
		E^n
		&\leq c_1(\triangle t)^2 \sum_{j=0}^{n-1}(1+c_2\triangle t)^j
		=\frac{c_1}{c_2}\left[(1+c_2\triangle t)^n-1\right]\triangle t\\
		&\leq\frac{c_1}{c_2}\left(\mathrm{e}^{c_2n\triangle t}-1\right)\triangle t
		\leq\frac{c_1}{c_2}\left(\mathrm{e}^{c_2T}-1\right)\triangle t
		\leq c\triangle t,
		\qquad n=0,1,\ldots,n_{\triangle t}.
	\end{aligned}
\end{equation}
The final time enters the constant through the interval~$\mathcal{I}$ and the factor $\mathrm{e}^{c_2T}$, and the index $n$ does not enter it. The quadratic error is thus bounded by a multiple of the time step size, and its source is the variance in Eq.~\eqref{eq:apx_continuous_limit_completed_square}, which the splitting of the weights produces in every step. As the quadratic error~\eqref{eq:apx_continuous_limit_quadratic_error} is the squared Wasserstein distance of order two~\cite{Villani2009} between the distribution and the Dirac measure $\delta_{d(t^n)}$ at the exact damage state, the distribution converges to this measure, uniformly in the time steps. The estimate bounds the effective degradation, in contrast, by the first absolute error moment, which the Cauchy--Schwarz inequality~\eqref{eq:apx_continuous_limit_first_error_moment} bounds by the square root of the quadratic error. This is the origin of the half order in the estimate~\eqref{eq:theory_scalar_model_continuous_limit_estimate}. The estimate is an upper bound. The mechanism which may render it sharp acts where the driving force stays close to the activation threshold, as the spread of the distribution may then enter the effective degradation, whereas both formulations converge at least at first order in the computations of Sec.~\ref{sec:numerics_unit_cell}. It remains to translate the bound~\eqref{eq:apx_continuous_limit_error_bound} into an estimate for the effective degradation. As in Eq.~\eqref{eq:apx_continuous_limit_new_degradation_error}, the reciprocals of two values not smaller than one differ by at most their difference. Together with the Lipschitz continuity of the reciprocal degradation on the interval~$\mathcal{I}$, Eqs.~\eqref{eq:apx_continuous_limit_effective_degradation},~\eqref{eq:apx_continuous_limit_first_error_moment} and~\eqref{eq:apx_continuous_limit_error_bound} lead to the estimate
\begin{equation}\label{eq:apx_continuous_limit_final_estimate}
	\begin{aligned}
		\left|\bar f_{\triangle t}^n - f(d(t^n))\right| &\leq \left|\sum_{k=0}^n \rho_k^n g(\hat d_k) - g(d(t^n))\right| \leq \sum_{k=0}^n \rho_k^n \left|g(\hat d_k) - g(d(t^n))\right|\\
		&\leq c \sum_{k=0}^n \rho_k^n |\hat d_k-d(t^n)| \leq c\sqrt{E^n} \leq c\sqrt{\triangle t}, \qquad n=0,1,\ldots,n_{\triangle t}.
	\end{aligned}
\end{equation}
The piecewise constant interpolation announced at the beginning of this section is defined by
\begin{equation}\label{eq:apx_continuous_limit_interpolation}
	\bar f_{\triangle t}(t) := \bar f_{\triangle t}^n, \qquad t \in [t^n,t^{n+1}) \quad\text{and}\quad n=0,1,\ldots,n_{\triangle t}-1, \qquad \bar f_{\triangle t}(T) := \bar f_{\triangle t}^{n_{\triangle t}}.
\end{equation}
The interpolation~\eqref{eq:apx_continuous_limit_interpolation} takes the nodal value at every node of the time grid. Hence, the estimate~\eqref{eq:apx_continuous_limit_final_estimate} proves Eq.~\eqref{eq:theory_scalar_model_continuous_limit_estimate}. Within the time step from time $t^n$ to time $t^{n+1}$, the degradation along the exact solution is Lipschitz continuous in time, as the degradation function is Lipschitz continuous on the interval~$\mathcal{I}$ and the damage rate is bounded by the damage-rate limitation. Together with the estimate~\eqref{eq:apx_continuous_limit_final_estimate} and as the time step size does not exceed the final time, this implies
\begin{equation}\label{eq:apx_continuous_limit_uniform_estimate}
	\begin{aligned}
		\left|\bar f_{\triangle t}(t) - f(d(t))\right| &\leq \left|\bar f_{\triangle t}^n - f(d(t^n))\right| + \left|f(d(t^n))-f(d(t))\right|\\
		&\leq c\sqrt{\triangle t} + c\triangle t \leq c\sqrt{\triangle t}, \qquad t \in [t^n,t^{n+1}).
	\end{aligned}
\end{equation}
For the final time $t=T$, the estimate~\eqref{eq:apx_continuous_limit_final_estimate} with $n=n_{\triangle t}$ applies directly, by the interpolation~\eqref{eq:apx_continuous_limit_interpolation}. Taking the supremum over the time interval $[0,T]$ proves the uniform convergence~\eqref{eq:theory_scalar_model_continuous_limit}.

\subsection{Convergence of the frozen-metric discretization}
\label{apx:frozen_metric_convergence}

We show that the frozen-metric discretization converges at first order. Under the assumptions of Appendix~\ref{apx:continuous_limit}, let the frozen-metric update~\eqref{eq:theory_frozen_metric_update} generate the states $z_n$ on the time grid~\eqref{eq:theory_scalar_model_time_grid} under the strains $\feps(t^n)$, with the damage states
\begin{equation} \label{eq:theory_frozen_metric_convergence_initial_state}
	z_0 := g(d_0) \quad\text{and}\quad d_n := g^{-1}(z_n), \qquad n=0,1,\ldots,n_{\triangle t}.
\end{equation}
Then, there is a constant $c>0$, independent of the number of load steps $n_{\triangle t}$, such that
\begin{equation} \label{eq:theory_frozen_metric_convergence_estimate}
	\max_{0\leq n\leq n_{\triangle t}} \left| f(d_n)-f(d(t^n)) \right| + \max_{0\leq n\leq n_{\triangle t}} \left| d_n-d(t^n) \right| \leq c\,\triangle t
\end{equation}
The constant $c$ grows exponentially with the Lipschitz constant $\ddotmax/\lambda$ of the kinetic function and thus deteriorate as the entropy parameter $\lambda$ tends to zero.

We write the update as a recursion for the reciprocal degradation, bound its defect at the exact states and accumulate the defects over the time steps. Throughout, $c$ denotes a positive constant independent of the number of load steps $n_{\triangle t}$, whose value may change from one occurrence to the next.

We start with the discrete evolution law. The damage rate depends on the driving force through the kinetic function~\eqref{eq:apx_continuous_limit_kinetic_function} of Appendix~\ref{apx:continuous_limit}, which is continuous, nondecreasing, bounded by the damage-rate limitation $\ddotmax$, vanishes up to the transition point $r+\lambda$ and is Lipschitz continuous by the estimate~\eqref{eq:apx_continuous_limit_kinetic_lipschitz}. On the active branch, the derivative~\eqref{eq:theory_scalar_model_dissipation_derivatives} of the dissipation potential turns the scalar equation~\eqref{eq:theory_frozen_metric_scalar_equation} into an equation for the rate variable $s$,

\begin{equation}\label{eq:apx_frozen_metric_convergence_rate_equation}
	r + \frac{\lambda}{1-s/\ddotmax} = \frac{m(z_n)\,\freeEnergy_0(\feps)}{z^2}, \qquad s := \frac{z-z_n}{\triangle t\,m(z_n)},
\end{equation}
which solves as
\begin{equation}\label{eq:apx_frozen_metric_convergence_rate}
	s = K\!\left(\frac{m(z_n)\,\freeEnergy_0(\feps)}{z^2}\right).
\end{equation}
On the inactive branch~\eqref{eq:theory_frozen_metric_inactive_condition}, the state remains at the previous state $z_n$, at which the argument of the kinetic function does not exceed the transition point. Under the strain $\feps(t^{n+1})$, both branches of the update~\eqref{eq:theory_frozen_metric_update} thus collapse into the recursion

\begin{equation}\label{eq:apx_frozen_metric_convergence_discrete_law}
	z_{n+1} = z_n + \triangle t\,\widehat F\big(t^{n+1},z_n,z_{n+1}\big)
\end{equation}
with the function
\begin{equation}\label{eq:apx_frozen_metric_convergence_rate_function}
	\widehat F(t,\eta,z) := m(\eta)\,K\!\left(\frac{m(\eta)\,\freeEnergy_0(\feps(t))}{z^2}\right), \qquad t\in[0,T] \quad\text{and}\quad \eta,z\geq 1 .
\end{equation}
The exact reciprocal degradation $z(t):=g(d(t))$ obeys the evolution
\begin{equation}\label{eq:apx_frozen_metric_convergence_exact_law}
	\dot z(t) = \widehat F\big(t,z(t),z(t)\big), \qquad t\in[0,T],
\end{equation}
which follows from the evolution law~\eqref{eq:theory_scalar_model_evolution} by the chain rule~\eqref{eq:theory_hidden_convexity_rate_transformation_preliminary}, as the derivative of the reciprocal degradation at the damage state is the metric~\eqref{eq:theory_hidden_convexity_metric} at the reciprocal degradation state.

The rate bounds~\eqref{eq:theory_scalar_model_rate_bound} and~\eqref{eq:theory_frozen_metric_physical_rate_bound} confine the exact and the discrete states to the interval
\begin{equation}\label{eq:apx_frozen_metric_convergence_range}
	\mathcal{Z} := \left[1,\,g(d_0+T\ddotmax)\right]
\end{equation}
for every number of load steps, which is compact by the condition~\eqref{eq:theory_scalar_model_attainable_damage_range}. On this interval, the metric is continuously differentiable by the identities~\eqref{eq:theory_hidden_convexity_inverse_derivatives} and the assumptions~\eqref{eq:theory_scalar_model_regularities}, and the undamaged energy along the strain path~\eqref{eq:theory_scalar_model_strain_path_regularity} is Lipschitz continuous in time. The function $\widehat F$ is thus Lipschitz continuous on the set $[0,T]\times\mathcal Z\times\mathcal Z$, with a constant $L$ that inherits the factor $1/\lambda$ of the kinetic function. Moreover, the function $\widehat F$ is bounded and nonincreasing in its third argument, as the kinetic function is bounded and nondecreasing and its argument $m(\eta)\,\freeEnergy_0(\feps(t))/z^2$ does not increase in the third argument $z$, the metric being positive and the undamaged energy nonnegative. By the exact evolution~\eqref{eq:apx_frozen_metric_convergence_exact_law}, the exact state and its rate are thus Lipschitz continuous in time, despite the kink of the kinetic function, the rate as the composition of the Lipschitz function $\widehat F$ with the Lipschitz path $t\mapsto(t,z(t),z(t))$.

Inserting the exact states into the recursion, the defect splits into the defect of the implicit Euler step and the change of the metric over the time step,
\begin{equation}\label{eq:apx_frozen_metric_convergence_defect}
	\begin{aligned}
		R_{n+1} &:= z(t^{n+1}) - z(t^n) - \triangle t\,\widehat F\big(t^{n+1},z(t^n),z(t^{n+1})\big)\\
		&= \int_{t^n}^{t^{n+1}}\big[\dot z(\tau)-\dot z(t^{n+1})\big]\,\diff\tau + \triangle t\,\Big[\widehat F\big(t^{n+1},z(t^{n+1}),z(t^{n+1})\big)-\widehat F\big(t^{n+1},z(t^n),z(t^{n+1})\big)\Big].
	\end{aligned}
\end{equation}
The Lipschitz continuity of the exact rate bounds the integrand by the product $c\,\triangle t$, and that of the function $\widehat F$ and of the exact state bounds the second bracket in the same way. Both contributions, and hence the defect, are thus of second order in the time step size,
\begin{equation}\label{eq:apx_frozen_metric_convergence_defect_bound}
	|R_{n+1}| \leq \int_{t^n}^{t^{n+1}} c\,\triangle t\,\diff\tau + \triangle t\,L\,\big|z(t^{n+1})-z(t^n)\big| \leq c\,(\triangle t)^2, \qquad n=0,1,\ldots,n_{\triangle t}-1.
\end{equation}

Subtracting the defect from the recursion and inserting the function $\widehat F$ with the metric at the discrete previous state and the exact new state, $\widehat F(t^{n+1},z_n,z(t^{n+1}))$, the error
\begin{equation}\label{eq:apx_frozen_metric_convergence_error}
	e_n:=z_n-z(t^n), \qquad n=0,1,\ldots,n_{\triangle t},
\end{equation}
satisfies

\begin{equation}\label{eq:apx_frozen_metric_convergence_error_identity}
	e_{n+1} - \triangle t\,D_n = e_n + \triangle t\,G_n - R_{n+1}
\end{equation}
with the differences
\begin{equation}\label{eq:apx_frozen_metric_convergence_differences}
	\begin{aligned}
		D_n &:= \widehat F\big(t^{n+1},z_n,z_{n+1}\big)-\widehat F\big(t^{n+1},z_n,z(t^{n+1})\big),\\
		G_n &:= \widehat F\big(t^{n+1},z_n,z(t^{n+1})\big)-\widehat F\big(t^{n+1},z(t^n),z(t^{n+1})\big).
	\end{aligned}
\end{equation}
The difference $D_n$ compares the new states in the third argument, in which the function $\widehat F$ does not increase. It thus vanishes or carries the sign opposite to the new error, so that the magnitudes on the left add, without any restriction on the time step size. The difference $G_n$ compares the old states and is bounded by the product $L\,|e_n|$ of the Lipschitz constant and the old error. Taking the magnitude of the error identity~\eqref{eq:apx_frozen_metric_convergence_error_identity}, applying the triangle inequality on the right and inserting the defect bound~\eqref{eq:apx_frozen_metric_convergence_defect_bound} leads to the recursion
\begin{equation}\label{eq:apx_frozen_metric_convergence_recursion}
	\begin{aligned}
		|e_{n+1}| &\leq |e_{n+1}| + \triangle t\,|D_n| = |e_{n+1}-\triangle t\,D_n| = |e_n + \triangle t\,G_n - R_{n+1}|\\
		&\leq |e_n| + \triangle t\,|G_n| + |R_{n+1}| \leq (1+L\triangle t)\,|e_n| + c\,(\triangle t)^2, \qquad n=0,1,\ldots,n_{\triangle t}-1,
	\end{aligned}
\end{equation}
whose first equality is the addition of the magnitudes on the left. The initial error vanishes by the initial state~\eqref{eq:theory_frozen_metric_convergence_initial_state}. Raising the elementary inequality
\begin{equation}\label{eq:apx_frozen_metric_convergence_exp_inequality}
	1+x \leq \mathrm{e}^x, \qquad x\geq 0,
\end{equation}
for the product $L\triangle t$ to the power of the summation index $j$ and bounding the products $j\triangle t$ and $n\triangle t$ by the final time $T$, we arrive at
\begin{equation}\label{eq:apx_frozen_metric_convergence_error_bound}
	\begin{aligned}
		|e_n| &\leq (1+L\triangle t)^n\,|e_0| + c\,(\triangle t)^2\sum_{j=0}^{n-1}(1+L\triangle t)^j \leq c\,(\triangle t)^2\sum_{j=0}^{n-1}\mathrm{e}^{L\,j\triangle t}\\
		&\leq c\,n\,(\triangle t)^2\,\mathrm{e}^{LT} \leq c\,\triangle t, \qquad n=0,1,\ldots,n_{\triangle t}.
	\end{aligned}
\end{equation}
The defects of second order thus accumulate over a number of steps that grows inversely with the time step size, which leaves an error of first order.

Last, the degradation is the reciprocal~\eqref{eq:theory_scalar_model_reciprocal_degradation} of the state, with Lipschitz constant one on the interval~$\mathcal Z$, and the derivative~\eqref{eq:theory_hidden_convexity_inverse_derivatives} of the inverse function is bounded by the reciprocal of the initial slope $g'(0)$, as the reciprocal degradation is convex and the damage states are nonnegative. Hence, bounding each term separately and taking the maxima over the index $n$,
\begin{equation}\label{eq:apx_frozen_metric_convergence_back}
	\big|f(d_n)-f(d(t^n))\big| + \big|d_n-d(t^n)\big| \leq \left(1+\frac{1}{g'(0)}\right)|e_n| \leq c\,\triangle t, \qquad n=0,1,\ldots,n_{\triangle t},
\end{equation}
and the estimate~\eqref{eq:theory_frozen_metric_convergence_estimate} follows for every positive integer $n_{\triangle t}$.

\section{Arguments for the frozen-metric discretization}
\label{apx:frozen_metric_properties}

\subsection{Joint convexity and constitutive update}
\label{apx:frozen_metric_convexity}

In this section, we derive the joint convexity of the frozen-metric increment. We recall the incremental potential~\eqref{eq:theory_frozen_metric_incremental_potential_explicit} in the form used below,
\begin{equation} \label{eq:apx_frozen_metric_potential}
	\freeEnergy_{\mathrm{inc}}(\feps,z\,| z_n) = \frac{1}{2z}\,\feps:\C:\feps + \triangle t\,\dissipot(s), \qquad s := \frac{z-z_n}{\triangle t\,m(z_n)},
\end{equation}
where the previous reciprocal degradation state $z_n\in[1,\infty)$ and the positive time step size $\triangle t$ are fixed throughout. The effective domain of the dissipation potential~\eqref{eq:theory_scalar_model_dissipation} confines the rate variable $s$ to the half-open interval $[0,\ddotmax)$, which is the admissibility condition~\eqref{eq:theory_frozen_metric_admissibility_preliminary}. At a vanishing rate, the derivatives of the dissipation potential are understood as right-sided limits, which extend the identities~\eqref{eq:theory_scalar_model_dissipation_derivatives} continuously to a vanishing damage rate. We start with the joint convexity. Differentiating the increment~\eqref{eq:apx_frozen_metric_potential} twice, and taking into account that the rate variable $s$ is affine in the reciprocal degradation state, yields the entries
\begin{equation} \label{eq:apx_frozen_metric_hessian}
	\frac{\partial^2\freeEnergy_{\mathrm{inc}}}{\partial\feps\partial\feps} = \frac{\C}{z}, \qquad \frac{\partial^2\freeEnergy_{\mathrm{inc}}}{\partial\feps\partial z} = -\frac{\C:\feps}{z^2} \qquad\text{and}\qquad \frac{\partial^2\freeEnergy_{\mathrm{inc}}}{\partial z^2} = \frac{\feps:\C:\feps}{z^3} + \frac{\dissipot''(s)}{\triangle t\,[m(z_n)]^2}
\end{equation}
of its Hessian. For a strain variation $\fxi\in\Sym{3}$ and a variation $\delta z\in\R$ of the reciprocal degradation state, completing the square in the entries~\eqref{eq:apx_frozen_metric_hessian} leads to the identity
\begin{equation} \label{eq:apx_frozen_metric_quadratic_form}
	\begin{aligned}
		&\frac{1}{z}\,\fxi:\C:\fxi - \frac{2\delta z}{z^2}\,\fxi:\C:\feps + \frac{(\delta z)^2}{z^3}\,\feps:\C:\feps + \frac{\dissipot''(s)}{\triangle t\,[m(z_n)]^2}\,(\delta z)^2 \\
		&\qquad = \frac{1}{z}\left(\fxi-\frac{\delta z}{z}\feps\right):\C:\left(\fxi-\frac{\delta z}{z}\feps\right) + \frac{\dissipot''(s)}{\triangle t\,[m(z_n)]^2}\,(\delta z)^2 .
	\end{aligned}
\end{equation}
Both terms on the right-hand side of Eq.~\eqref{eq:apx_frozen_metric_quadratic_form} are nonnegative, because the elasticity tensor $\C$ is positive definite and the dissipation potential is strictly convex on its effective domain, see Eq.~\eqref{eq:theory_scalar_model_dissipation_derivatives}. The second term vanishes only for a vanishing state variation, and the first is then positive unless the strain variation vanishes as well. Thus, the frozen-metric increment is strictly convex in the strain and the reciprocal degradation state jointly on the admissible set~\eqref{eq:theory_frozen_metric_admissibility}. The logarithmic term of the dissipation potential~\eqref{eq:theory_scalar_model_dissipation} moreover diverges as the rate variable $s$ approaches the damage-rate limitation $\ddotmax$, so that the increment~\eqref{eq:apx_frozen_metric_potential} tends to infinity at the upper endpoint of the admissible interval and the update~\eqref{eq:theory_frozen_metric_update} attains its minimum. Hence, the update is a well-defined single-valued function of the strain. The stress~\eqref{eq:theory_frozen_metric_stress} is the partial derivative of the increment~\eqref{eq:apx_frozen_metric_potential} with respect to the strain at a fixed reciprocal degradation state. It is at the same time the stress carried by the incremental potential condensed over the reciprocal degradation state,
\begin{equation} \label{eq:apx_frozen_metric_condensed}
	\freeEnergy_{\mathrm{cond}}(\feps| z_n) := \min\left\{\freeEnergy_{\mathrm{inc}}(\feps,\zeta| z_n) \,\middle|\, z_n \leq \zeta < z_n + m(z_n)\triangle t\,\ddotmax\right\},
\end{equation}
whose derivative reads
\begin{equation} \label{eq:apx_frozen_metric_envelope}
	\frac{\partial\freeEnergy_{\mathrm{cond}}}{\partial\feps} = \frac{\partial\freeEnergy_{\mathrm{inc}}}{\partial\feps} + \frac{\partial\freeEnergy_{\mathrm{inc}}}{\partial z}\,\frac{\partial z}{\partial\feps} = \frac{1}{z}\,\C:\feps .
\end{equation}
Indeed, the middle term drops out because the reciprocal degradation state does not respond to the strain on the inactive branch, whereas the derivative of the increment with respect to the state vanishes on the active branch by the stationarity of the update~\eqref{eq:theory_frozen_metric_update}. On the switching surface between both branches, where the inactive condition~\eqref{eq:theory_frozen_metric_inactive_condition} holds with equality, the state is not differentiable in the strain. It is continuous, however, as the previous state solves the stationarity condition~\eqref{eq:apx_frozen_metric_stationarity} there, so that the derivative~\eqref{eq:apx_frozen_metric_envelope} extends continuously across this surface. Stress and consistent algorithmic material tangent therefore both descend from the single potential~\eqref{eq:apx_frozen_metric_condensed}, which is continuously differentiable and whose second derivative jumps across the switching surface. The update~\eqref{eq:theory_frozen_metric_update} is decided by the derivative of the increment with respect to the reciprocal degradation state,
\begin{equation} \label{eq:apx_frozen_metric_stationarity}
	\frac{\partial\freeEnergy_{\mathrm{inc}}}{\partial z}(\feps,z\,| z_n) = -\frac{\freeEnergy_0(\feps)}{z^2} + \frac{\dissipot'(s)}{m(z_n)} .
\end{equation}
As the increment is convex in the state, the previous state minimizes precisely if the derivative~\eqref{eq:apx_frozen_metric_stationarity} is nonnegative there, and evaluating the first identity in Eq.~\eqref{eq:theory_scalar_model_dissipation_derivatives} at a vanishing rate turns this requirement into the inactive condition~\eqref{eq:theory_frozen_metric_inactive_condition}. Otherwise, the minimizer lies in the interior of the admissible interval, the derivative~\eqref{eq:apx_frozen_metric_stationarity} vanishes there, and the scalar equation~\eqref{eq:theory_frozen_metric_scalar_equation} follows.

\subsection{Pole-free residual and monotone Newton iteration}
\label{apx:frozen_metric_Newton}

In this section, we show that the undamped Newton iteration for the pole-free residual~\eqref{eq:theory_frozen_metric_pole_free_residual}, started at the previous state, converges monotonically to the active root. Taking into account the explicit derivative of the dissipation potential in Eq.~\eqref{eq:theory_scalar_model_dissipation_derivatives}, the scalar stationarity equation~\eqref{eq:theory_frozen_metric_scalar_equation} reads
\begin{equation}
\label{eq:apx_frozen_metric_scalar_equation_explicit}
	r + \frac{\lambda}{1-\dfrac{z-z_n}{\triangle t\,m(z_n)\,\ddotmax}} = m(z_n)\frac{\freeEnergy_0(\feps)}{z^2}.
\end{equation}
Throughout this section the update is active, that is, the inactive condition~\eqref{eq:theory_frozen_metric_inactive_condition} is violated and the abbreviations~\eqref{eq:theory_frozen_metric_scalar_equation_abbreviations} obey
\begin{equation}\label{eq:apx_frozen_metric_active_condition}
	A_n>z_n^2(r+\lambda)>0.
\end{equation}
The residual associated with the scalar equation~\eqref{eq:apx_frozen_metric_scalar_equation_explicit},
\begin{equation}
\label{eq:apx_frozen_metric_residual}
	P(z) := r + \frac{\lambda}{1-\dfrac{z-z_n}{\alpha_n}} - \frac{A_n}{z^2},
\end{equation}
carries a simple pole at the upper endpoint of the admissible interval~\eqref{eq:theory_frozen_metric_admissibility}, where the affine factor in its denominator, the relative distance of the transformed damage rate from the damage-rate limitation, vanishes. The active minimizer lies strictly inside the admissible interval, so that multiplication by this positive factor removes the pole without changing the active root and leads to the pole-free residual~\eqref{eq:theory_frozen_metric_pole_free_residual}. The first two derivatives of the pole-free residual satisfy the identities
\begin{equation}
\label{eq:apx_frozen_metric_pole_free_residual_derivative}
	M'(z) = \frac{2A_n}{z^3}\left(1-\frac{z-z_n}{\alpha_n}\right) - \frac{1}{\alpha_n}\left(r-\frac{A_n}{z^2}\right)
\end{equation}
and
\begin{equation}
\label{eq:apx_frozen_metric_pole_free_residual_second_derivative}
	M''(z) = -\frac{4A_n}{\alpha_n z^3} - \frac{6A_n}{z^4}\left(1-\frac{z-z_n}{\alpha_n}\right) < 0, \qquad z_n \leq z \leq z_n+\alpha_n.
\end{equation}
The pole-free residual is therefore strictly concave, and it takes the boundary values
\begin{equation}
\label{eq:apx_frozen_metric_pole_free_boundary_values}
	M(z_n) = r+\lambda-\frac{A_n}{z_n^2} < 0 \quad\text{and}\quad M(z_n+\alpha_n) = \lambda > 0.
\end{equation}
The sign change and the continuity of the function $M$ provide a root $z_\ast$ in the interior of the admissible interval, and the strict concavity renders it unique, because beyond the larger of two roots a strictly concave function is negative, which the positive value at the upper endpoint excludes. The root coincides with the unique active minimizer of the incremental problem~\eqref{eq:theory_frozen_metric_update}. Moreover, the derivative satisfies
\begin{equation}
\label{eq:apx_frozen_metric_pole_free_derivative_positive}
	M'(z) > 0, \qquad z_n \leq z \leq z_\ast.
\end{equation}
Indeed, a nonpositive derivative at the root would, by the concavity of the function $M$, keep the function nonpositive to the right of the root, contradicting the positive boundary value in Eq.~\eqref{eq:apx_frozen_metric_pole_free_boundary_values}. As the derivative $M'$ is strictly decreasing, its positivity at the root extends to the entire interval from the previous state $z_n$ to the active root $z_\ast$. The previous reciprocal degradation state lies strictly below the active root and therefore initializes the undamped Newton iteration
\begin{equation}
\label{eq:apx_frozen_metric_Newton_iteration}
	z^{(k+1)} := z^{(k)} - \frac{M(z^{(k)})}{M'(z^{(k)})}, \qquad z^{(0)} := z_n.
\end{equation}
At an iterate with $z_n\leq z^{(k)}<z_\ast$, the residual is negative, whereas its derivative is positive by Eq.~\eqref{eq:apx_frozen_metric_pole_free_derivative_positive}. Consequently, the Newton increment is positive and
\begin{equation}
\label{eq:apx_frozen_metric_Newton_lower_bound}
	z^{(k+1)} > z^{(k)}.
\end{equation}
As the graph of the concave function $M$ lies below its tangent at the iterate $z^{(k)}$, evaluating the tangent inequality at the root leads to
\begin{equation}
\label{eq:apx_frozen_metric_Newton_tangent_bound}
	0 = M(z_\ast) \leq M(z^{(k)}) + M'(z^{(k)})\left(z_\ast-z^{(k)}\right),
\end{equation}
which, the derivative being positive, is equivalent to
\begin{equation}
\label{eq:apx_frozen_metric_Newton_upper_bound}
	z^{(k+1)} = z^{(k)} - \frac{M(z^{(k)})}{M'(z^{(k)})} \leq z_\ast.
\end{equation}
The inequalities~\eqref{eq:apx_frozen_metric_Newton_lower_bound} and~\eqref{eq:apx_frozen_metric_Newton_upper_bound} show inductively~\cite{Ortega1970} that

\begin{equation}
\label{eq:apx_frozen_metric_Newton_invariant}
	z_n = z^{(0)} \leq z^{(1)} \leq \ldots \leq z^{(k)} \leq z_\ast < z_n+\alpha_n,
\end{equation}
where $z^{(j)}<z^{(j+1)}$ holds as long as $z^{(j)}<z_\ast$, in particular for $j=0$.
Thus, every Newton iterate remains in the admissible interval, and the sequence converges monotonically from below to a limit $\bar z\leq z_\ast$. Passing to the limit in the Newton iteration~\eqref{eq:apx_frozen_metric_Newton_iteration} yields $M(\bar z)=0$, as the derivative $M'(\bar z)$ is positive by Eq.~\eqref{eq:apx_frozen_metric_pole_free_derivative_positive}, and the uniqueness of the active root implies $\bar z=z_\ast$.

\subsection{Consistent algorithmic material tangent}
\label{apx:frozen_metric_tangent}

We derive the consistent algorithmic material tangent on the active branch from the stress relation~\eqref{eq:theory_frozen_metric_stress}. As the reciprocal degradation state depends on the strain through the incremental minimization problem~\eqref{eq:theory_frozen_metric_update}, differentiation gives
\begin{equation}\label{eq:apx_frozen_metric_tangent_start}
	\C_{\mathrm{alg}} = \frac{\diff}{\diff\feps}\left[\frac{1}{z}\,\C:\feps\right] = \frac{1}{z}\left[\C-\frac{1}{z}(\C:\feps)\otimes\frac{\partial z}{\partial\feps}\right].
\end{equation}
The response of the reciprocal degradation state follows from the implicit function theorem applied to the stationarity condition, i.e.\ the vanishing of the derivative~\eqref{eq:apx_frozen_metric_stationarity}, as the curvature entry of the Hessian~\eqref{eq:apx_frozen_metric_hessian} is positive,
\begin{equation}\label{eq:apx_frozen_metric_state_response}
	\frac{\partial z}{\partial\feps} = -\left[\frac{\partial^2\freeEnergy_{\mathrm{inc}}}{\partial z^2}\right]^{-1}\frac{\partial^2\freeEnergy_{\mathrm{inc}}}{\partial\feps\partial z} = \frac{\C:\feps}{z^2\left[\dfrac{\feps:\C:\feps}{z^3}+\dfrac{\dissipot''(s)}{\triangle t\,[m(z_n)]^2}\right]},
\end{equation}
where the rate variable $s$ is defined in Eq.~\eqref{eq:apx_frozen_metric_potential}. Inserting the state response~\eqref{eq:apx_frozen_metric_state_response} into Eq.~\eqref{eq:apx_frozen_metric_tangent_start} leads to the rank-one representation
\begin{equation}\label{eq:apx_frozen_metric_tangent_rank_one}
	\C_{\mathrm{alg}} = \frac{1}{z}\left[\C-\kappa(\C:\feps)\otimes(\C:\feps)\right],
\end{equation}
where
\begin{equation}\label{eq:apx_frozen_metric_rank_one_coefficient_preliminary}
	\kappa := \frac{1}{z^3}\left[\frac{\feps:\C:\feps}{z^3}+\frac{\dissipot''(s)}{\triangle t\,[m(z_n)]^2}\right]^{-1}.
\end{equation}
Taking into account the second identity in Eq.~\eqref{eq:theory_scalar_model_dissipation_derivatives} and the definition~\eqref{eq:theory_frozen_metric_tangent_factor}, the dissipative curvature satisfies
\begin{equation}\label{eq:apx_frozen_metric_dissipative_curvature}
	\frac{\dissipot''(s)}{\triangle t\,[m(z_n)]^2} = \frac{\lambda}{\triangle t\,\ddotmax\,\gamma}.
\end{equation}
Consequently, the rank-one coefficient~\eqref{eq:apx_frozen_metric_rank_one_coefficient_preliminary} becomes
\begin{equation}\label{eq:apx_frozen_metric_rank_one_coefficient}
	\kappa = \frac{\triangle t\,\ddotmax\,\gamma}{\lambda z^3+\triangle t\,\ddotmax\,(\feps:\C:\feps)\gamma},
\end{equation}
and the rank-one representation~\eqref{eq:apx_frozen_metric_tangent_rank_one} turns into the consistent algorithmic material tangent~\eqref{eq:theory_frozen_metric_material_tangent}. Built from the single symmetric second-order tensor $\C:\feps$, the rank-one correction in Eq.~\eqref{eq:apx_frozen_metric_tangent_rank_one} leaves the consistent algorithmic material tangent symmetric. For a strain variation $\fxi\in\Sym{3}$, the Cauchy--Schwarz inequality with respect to the scalar product induced by the positive-definite elasticity tensor $\C$,
\begin{equation}\label{eq:apx_frozen_metric_Cauchy_Schwarz}
	(\fxi:\C:\feps)^2 \leq (\fxi:\C:\fxi)(\feps:\C:\feps),
\end{equation}
combines with the tangent~\eqref{eq:theory_frozen_metric_material_tangent} into the lower bound
\begin{equation}\label{eq:apx_frozen_metric_tangent_lower_bound}
	\fxi:\C_{\mathrm{alg}}:\fxi \geq \frac{1}{z}\frac{\lambda z^3}{\lambda z^3+\triangle t\,\ddotmax\,(\feps:\C:\feps)\gamma}\,\fxi:\C:\fxi,
\end{equation}
whose right-hand side is positive for every non-vanishing strain variation. Hence, the consistent algorithmic material tangent is positive definite on the active branch. The lower bound~\eqref{eq:apx_frozen_metric_tangent_lower_bound} still depends on the current strain, a dependence which the active stationarity equation removes. The remaining rate fraction
\begin{equation}\label{eq:apx_frozen_metric_remaining_rate_fraction}
	q:=1-\frac{z-z_n}{\alpha_n}
\end{equation}
obeys $0<q<1$ on the active branch. The definitions~\eqref{eq:theory_frozen_metric_scalar_equation_abbreviations} and~\eqref{eq:theory_frozen_metric_tangent_factor} give
\begin{equation}\label{eq:apx_frozen_metric_rate_fraction_identities}
	z=z_n+\alpha_n(1-q) \quad\text{and}\quad \gamma=[m(z_n)]^2q^2,
\end{equation}
and the stationarity equation~\eqref{eq:apx_frozen_metric_scalar_equation_explicit} implies
\begin{equation}\label{eq:apx_frozen_metric_active_energy_identity}
	\feps:\C:\feps=\frac{2z^2}{m(z_n)}\left(r+\frac{\lambda}{q}\right).
\end{equation}
Inserting the identities~\eqref{eq:apx_frozen_metric_rate_fraction_identities} and~\eqref{eq:apx_frozen_metric_active_energy_identity} into the lower bound~\eqref{eq:apx_frozen_metric_tangent_lower_bound} leads to
\begin{equation}\label{eq:apx_frozen_metric_tangent_lower_bound_reduced}
	\fxi:\C_{\mathrm{alg}}:\fxi \geq \frac{\lambda}{\lambda z+2\alpha_n(rq^2+\lambda q)}\,\fxi:\C:\fxi,
\end{equation}
whose denominator satisfies
\begin{equation}\label{eq:apx_frozen_metric_tangent_denominator_bound}
	\lambda z+2\alpha_n(rq^2+\lambda q)=\lambda z_n+\alpha_n\left[\lambda(1+q)+2rq^2\right]\leq\lambda z_n+2\alpha_n(r+\lambda)
\end{equation}
by the first identity in Eq.~\eqref{eq:apx_frozen_metric_rate_fraction_identities}. Consequently, the uniform estimate~\eqref{eq:theory_frozen_metric_uniform_tangent_bound}
holds throughout the active branch. On the inactive branch, the reciprocal degradation state remains fixed at $z_n$, and direct differentiation of the stress~\eqref{eq:theory_frozen_metric_stress} gives the damaged elastic tangent~\eqref{eq:theory_frozen_metric_elastic_tangent}. As
\begin{equation}\label{eq:apx_frozen_metric_inactive_tangent_comparison}
	\frac{\lambda}{\lambda z_n+2\alpha_n(r+\lambda)}\leq\frac{1}{z_n},
\end{equation}
the lower bound~\eqref{eq:theory_frozen_metric_uniform_tangent_bound} also holds on the inactive branch. If $c_{\C}>0$ denotes the smallest eigenvalue of the undamaged elasticity tensor with respect to the Euclidean norm on $\Sym{3}$, then
\begin{equation}\label{eq:apx_frozen_metric_strong_convexity_bound}
	\fxi:\C_{\mathrm{alg}}:\fxi \geq \mu_n\|\fxi\|^2, \qquad \mu_n:=\frac{\lambda c_{\C}}{\lambda z_n+2\alpha_n(r+\lambda)}>0.
\end{equation}
The consistent algorithmic material tangent is therefore uniformly positive definite with respect to the current strain or, equivalently, the incremental potential condensed over the reciprocal degradation state is $\mu_n$-strongly convex in the strain. The equivalence persists although the second derivative jumps across the switching surface. By the inactive condition~\eqref{eq:theory_frozen_metric_inactive_condition}, this surface is a level set of the undamaged energy and thus meets every segment of strains at most twice, and the stress is continuous across it.

\bibliographystyle{ieeetr}

{\footnotesize\bibliography{literature}}

\end{document}